**Title:** One-sample aggregate data meta-analysis of medians
**Running title:** Meta-analysis of medians

**Authors:**

Sean McGrath[a,d]

XiaoFei Zhao[d]

Zhi Zhen Qin[e]

Russell Steele[a]

Andrea Benedetti[a,b,c,d,*†]

[a] Department of Mathematics and Statistics, McGill University, Montreal, Canada

[b] Department of Epidemiology, Biostatistics and Occupational Health, McGill University, Montreal, Canada

[c] Department of Medicine, McGill University, Montreal, Canada

[d] Respiratory Epidemiology and Clinical Research Unit (RECRU), McGill University Health Centre, Montreal, Canada

[e] Stop TB Partnership Secretariat, Geneva, Switzerland

* Correspondence to: Andrea Benedetti

† Email: andrea.benedetti@mcgill.ca

**Abstract**

An aggregate data meta-analysis is a statistical method that pools the summary statistics of several selected studies to estimate the outcome of interest. When considering a continuous outcome, typically each study must report the same measure of the outcome variable and its spread (e.g., the sample mean and its standard error). However, some studies may instead report the median along with various measures of spread. Recently, the task of incorporating medians in meta-analysis has been achieved by estimating the sample mean and its standard error from each study that reports a median in order to meta-analyze the means. In this paper, we propose two alternative approaches to meta-analyze data that instead rely on medians. We systematically compare these approaches via simulation study to each other and to methods that transform the study-specific medians and spread into sample means and their standard errors. We demonstrate that the proposed median-based approaches perform better than the transformation-based approaches, especially when applied to skewed data and data with high inter-study variance. In addition, when meta-analyzing data that consists of medians, we show that the median-based approaches perform considerably better than or comparably to the best-case scenario for a transformation approach: conducting a meta-analysis using the actual sample mean and standard error of the mean of each study. Finally, we illustrate these approaches in a meta-analysis of patient delay in tuberculosis diagnosis.

1. **Introduction**

Researchers increasingly use meta-analyses to synthesize conclusions across studies.[1] Most often, outcomes are assumed to follow a normal distribution or are transformed to follow a normal distribution. However, on occasion, the primary studies report a median. Such data typically arise from meta-analyses that evaluate time-based outcomes (e.g., patient delay). Asymptotically, the distribution of a median is normal with a variance that depends on the underlying outcome distribution. How to best meta-analyze data that consists of medians is currently not known.

In a typical one-sample meta-analysis of a continuous outcome, studies that investigate the outcome of interest are identified in the literature. Each study contributes summary measures of center and spread, often the sample mean its standard error. However, studies might provide a median with the first and third quartiles, minimum and maximum values, or other measures of spread.

This work was motivated by two studies that aimed to estimate various delays in patient populations.[2,3] Some studies reported a sample mean and its standard error, some reported a median with the minimum and maximum values, and others reported a median with the first and third quartiles. Presumably, if a median was reported, this was because the distribution was non-normal in the original data.

Several recently published papers propose methods of incorporating studies that report medians in meta-analyses.[4-7] Each of these methods assumes that the studies reporting medians present the sample size and one of the following measures of spread: the minimum and maximum values; the first and third quartiles; or, both. All of these methods estimate the sample mean and its standard error from each study that reports a median in order to perform a meta-analysis to estimate the study population mean.

In this work, we propose several approaches to meta-analyze medians and systematically compare them via simulation study to each other and to methods that transform study-specific medians and spread into sample means and standard their errors. In sensitivity analyses, we consider the case where some primary studies present means and their standard errors and others present medians along with first and third quartiles or minimum and maximum values, which occurs frequently in real-life data. Finally, we estimate pooled patient delay in tuberculosis diagnosis in China using our methods.

In the following section, we introduce the approaches to meta-analyze medians considered in the simulation study. We describe the design of the simulation study in Section 3. The results of the primary and sensitivity analyses are presented in Section 4. In Section 5, we apply the methods to an example data set to estimate pooled patient delay in tuberculosis diagnosis in China. We conclude with a discussion and guidelines for data analysts and authors in Section 6.

2. **Approaches to a One-Sample Meta-Analysis of Medians**

We review the inverse variance method that is typically used to pool studies in meta-analysis. For study $i$ where $i = 1, \ldots, k$, let $y_i$ denote the reported measure of effect and let $v_i$ denote its sampling variance. The pooled measure of effect is estimated by a weighted mean of the $y_i$ where the weight of study $i$ depends on its precision. That is, the estimated pooled measure of effect, $\hat{\mu}$, and its standard error are given by

$$\hat{\mu} = \frac{\sum_{i=1}^{k} y_i w_i}{\sum_{i=1}^{k} w_i}, \qquad \widehat{SE}(\hat{\mu}) = \sqrt{\frac{1}{\sum_{i=1}^{k} w_i}}$$

We use $w_i = v_i^{-1}$ in a fixed effect analysis and $w_i = (v_i + \hat{\tau}^2)^{-1}$ in a random effects analysis, where $\hat{\tau}^2$ is an estimate of between-study heterogeneity. Estimators of between-study heterogeneity are described by DerSimonian and Kacker[8] and Viechtbauer[9].

The challenge of applying the inverse variance method to pool studies that report the sample median is that the sampling variance of the median is unavailable. Individual studies do not typically report estimates of the sampling variance of the median, and estimating the sampling variance of the median from sample quantiles is challenging because it depends on the underlying distribution of the outcome.

Instead of using the median as the measure of effect in meta-analysis, several authors have proposed methods to estimate the sample mean and its sampling variance from studies that report medians[4-7]. These transformations are performed in order to pool the estimated means using the inverse variance method. We propose two alternative methods to pool the actual sample medians that are analogous to pooling studies in fixed effect and random effects models.

In this work, we consider approaches for meta-analyzing data from studies that report one of three sets of summary statistics: (i) the median, first and third quartiles, and number of subjects; (ii) the median, minimum and maximum values, and number of subjects; or, (iii) the sample mean and its standard error. The approaches we consider fall into two main categories: those that attempt to estimate a pooled mean, and those that attempt to estimate a pooled median. We use the following notation for the summary statistics that may be reported by a study: sample size ($n$), sample mean ($\bar{x}$), sample standard deviation ($s_x$), minimum value ($Q_{min}$), first quartile ($Q_1$), median ($Q_2$), third quartile ($Q_3$), and maximum value ($Q_{max}$).

*2.1. Approaches for estimating the study population mean*

**TRANSFORMATION 1 (T1)**

We use the method of Wan et al[5] to estimate the mean and its standard error for each study that reports a median, first and third quartiles, and number of subjects. The formulas for estimating the sample mean and standard deviation are as follows:

$$\bar{x} \approx \frac{Q_1 + Q_2 + Q_3}{3} \qquad (1)$$

$$s_x \approx \frac{Q_3 - Q_1}{2\Phi^{-1}\left(\frac{0.75n - 0.125}{n + 0.25}\right)} \qquad (2)$$

where $\Phi$ denotes the cumulative distribution function of the standard normal distribution. The standard error of the mean is obtained by dividing the estimated standard deviation by $\sqrt{n}$. Then, we estimate a fixed effect pooled mean and 95% confidence interval (CI) using the estimated means and standard errors of each study.[10] Additionally, using the method of DerSimonian and Laird[11] to estimate between-study heterogeneity, we estimate a random effects pooled mean and 95% CI from the estimated means and standard errors of each study. We denote the fixed effect method as T1$_{FE}$ and the random effects method as T1$_{RE}$.

To obtain these approximations, Wan et al[5] begin by assuming that the sample is drawn from a normal distribution with mean $\mu$ and standard deviation $\sigma$. The estimation of the sample mean follows from the observation that $E(Q_1 + Q_2 + Q_3) = 3\mu$. The sample mean is then estimated by $\frac{1}{3}E(Q_1 + Q_2 + Q_3)$ and replacing the expected value of the quartiles with the respective sample quartiles. For deriving the expression for the standard deviation, we introduce the following notation used by Wan et al[5]. Let $x_{(1)} \leq \cdots \leq x_{(n)}$ be the ordered observations, where $n = 4k + 1$ is assumed for simplicity. Moreover, let $Z_i \sim \text{Normal}(0,1)$ for $i = 1, \ldots, n$ and let $Z_{(1)}, \ldots, Z_{(n)}$ be the order statistics. Then $\mu + \sigma Z_{(1)}, \ldots, \mu + \sigma Z_{(n)}$ are the random variables corresponding to the ordered observations $x_{(1)}, \ldots, x_{(n)}$. It can be shown that $E(Q_3 - Q_1) = 2\sigma E(Z_{(3k+1)})$, and so we estimate the sample standard deviation by $E(Q_3 - Q_1)/[2E(Z_{(3k+1)})]$ and replace the expected value of the quantiles with the respective sample quartiles. When $n$ is large, $E(Z_{(3k+1)})$ can be approximated by $\Phi^{-1}\left(\frac{0.75n - 0.125}{n + 0.25}\right)$, yielding formula (2).

## TRANSFORMATION 2 (T2)

We apply the method of Hozo et al[5] to estimate the sample mean for each study that reports a median, minimum and maximum values, and number of subjects. This method estimates the sample mean by

$$\bar{x} \approx \frac{Q_{min} + 2Q_2 + Q_{max}}{4} \qquad (3)$$

The idea behind the approximation of Hozo et al[4] is as follows. Let $x_{(1)} \leq x_{(2)} \leq \cdots \leq x_{(n)}$ be the ordered sample of $n$ observations, where $n$ is assumed to be odd for simplicity. No distributional assumptions are made. We then have the following inequalities

$$Q_{min} \leq x_{(i)} \leq Q_2 \text{ for } i = 1, \ldots, \frac{n-1}{2}$$

$$Q_2 \leq x_{(i)} \leq Q_2 \text{ for } i = \frac{n+1}{2}$$

$$Q_2 \leq x_{(i)} \leq Q_{max} \text{ for } i = \frac{n+3}{2}, \ldots, n$$

Upon summing these $n$ inequalities and dividing by $n$, we obtain upper and lower bounds for the sample mean. We can estimate the sample mean by taking the average of the upper and lower bounds, which yields the following formula

$$\bar{x} \approx \frac{Q_{min} + 2Q_2 + Q_{max}}{4} + \frac{Q_{min} - 2Q_2 + Q_{max}}{4n}$$

Since the second term becomes negligible as $n$ becomes large, the sample mean estimate of Hozo et al[4] is obtained by removing the second term in the expression above.

We use the method of Wan et al[5] to estimate the standard error, which improves the method of Hozo et al[4] for estimating the standard deviation. This method estimates the standard deviation by

$$s_x \approx \frac{Q_{max} - Q_{min}}{2\Phi^{-1}\left(\frac{n-0.375}{n+0.25}\right)} \tag{4}$$

Wan et al[5] derived this expression under the same assumptions stated in the description of the T1 methods. Using the same notation introduced for the T1 method, it can be shown that $E(Q_{max} - Q_{min}) = 2\sigma E(Z_{(n)})$, and so we estimate the sample standard deviation $E(Q_{max} - Q_{min})/2E(Z_{(n)})$. We then estimate $E(Q_{min})$ and $E(Q_{max})$ with the sample minimum and maximum values, respectively. When $n$ is large, $E(Z_{(n)})$ can be approximated by $\Phi^{-1}\left(\frac{n-0.375}{n+0.25}\right)$, yielding formula (4).

After estimating the sample mean and standard deviation, we subsequently compute the standard error of the mean. We then estimate a fixed effect and random effects pooled mean and 95% CI,[10,11] denoted by T2$_{FE}$ and T2$_{RE}$, respectively.

**MEANS**

As a best case scenario for approaches estimating the mean, we consider using the sample mean and its standard error from each study. We estimate a fixed effect and random effects pooled mean and 95% confidence interval from the sample mean and its standard error of each study.[10,11] We denote the fixed effect method as MEANS$_{FE}$ and the random effects method as MEANS$_{RE}$ in this case. For the MEANS$_{RE}$ method, we estimate between-study heterogeneity using the method of DerSimonian and Laird[11].

*2.2. Approaches for estimating the study population median*

We propose the following two *median-based* approaches, as alternatives to transforming the study-specific medians and spread into sample means and their respective standard errors. These approaches do not require the studies to report a measure of spread, are non-parametric, and make no distributional assumption about the underlying distributions.

**Median of Medians (MM)**

In systematic reviews where individual studies report the median of the outcome, authors often report the median of the study-specific medians to summarize the data (e.g., see studies[2,12,13]). Therefore, we consider using the median of the study-specific medians as our pooled median estimate. We construct an approximate 95% CI around the median by using the $\frac{1}{2} - min\left\{\frac{1}{2}, \frac{z_{0.025}}{2\sqrt{k}}\right\}$ quantile of the study-specific medians as the lower limit and the $\frac{1}{2} + min\left\{\frac{1}{2}, \frac{z_{0.025}}{2\sqrt{k}}\right\}$ quantile as the upper limit, where $z_{0.025}$ is the 0.975 quantile of the standard normal distribution.[14] This method is akin random effects method with high estimated heterogeneity because each study is weighted equally.

**Weighted Median of Medians (WM)**

We propose a method analogous to the MM method that is appropriate for when a fixed effect analysis is desired. We use the weighted median of the study-specific medians as our pooled median estimate. For the approximate 95% CI, we use the $\frac{1}{2} - min\left\{\frac{1}{2}, \frac{z_{0.025}}{2\sqrt{k}}\right\}$ weighted quantile of the study specific medians as the lower limit and the $\frac{1}{2} + min\left\{\frac{1}{2}, \frac{z_{0.025}}{2\sqrt{k}}\right\}$ weighted quantile as the upper limit.[14] In a typical fixed effect analysis, studies are weighted by their precision, which we consider estimating by the sample size. That is, we use weights that are proportional to the number of subjects in the study and normalized to sum to 1.

The basic computation of a weighted quantile is as follows. Let $x_{(1)} \leq x_{(2)} \leq \cdots \leq x_{(n)}$ be an ordered sample of $n$ observations with corresponding weights $w_1, w_2, \ldots, w_n$ that are normalized to sum to 1. The weighted $q$ sample quantile can be estimated by the value $x_k$ where $k$ is the largest index such that $\sum_{i=1}^{k} w_i \leq q$. The computation of weighted quantiles was performed in R using the wtd.quantile function in the Hmisc package, which uses linear interpolation to estimate quantiles that are between two consecutive $x_{(i)}$ values.

### 3. Simulation Study

We performed a simulation study to systematically investigate various approaches of meta-analyzing medians.

*3.1. Data generation*

We simulated data typical of those collected in a one-sample aggregate data meta-analysis. We varied the number of studies in each meta-analysis to be either 15 or 50. The number of subjects in each study was generated from a log-normal distribution with median equal to 50 or 100, and scale parameter equal to 1. We eliminated very large or very small studies by the following: when the median number of subjects was 50, we excluded studies with less than 25 or more than 100 subjects; when the median number of subjects was 100, we excluded studies with less than 25 or more than 500 subjects.

We generated the outcome variable in each study from a log-normal distribution and varied the parameters to achieve distributions that were reasonably symmetric to substantially skewed. A random effect, generated from a normal distribution with mean 0 and variance $\tau^2$, was added to the location parameter of the log-normal distribution. Let the random effect $M$ be the study-specific median of the outcome variable of interest, $\tau^2$ be a measure of inter-study variance, and $\sigma^2$ be a measure of intra-study variance. Then, the following formula was used for generating outcomes:

$$M \sim \text{Normal}(0, \tau^2) \text{ for each meta-analysis,}$$

$$\text{Outcome} \sim m \times \text{Log-Normal}(M, \sigma^2) \text{ for each study in a meta-analysis.}$$

The following 13 combinations of $(\tau^2, \sigma^2)$ were used: (1/16, 1/16), (1/16, 1/4), (1/4, 1/16), (1/4, 1/4), (1/4, 1), (1, 1/16), (1, 1/4), (1, 1), (1, 4), (4, 1/16), (4, 1/4), (4, 1), and (4, 4). Some combinations of $(\tau^2, \sigma^2)$, such as (1/16, 4), are unlikely to correspond to any real-life scenario because $\tau^2$ and $\sigma^2$ differ by more than one order of magnitude, and so they were not used.

The data generation was performed in two steps. In the first step, $m$ assumed a value such that the expected value of the mean of outcomes was 5. In the second step, the value of $m$ was such that the expected value of the median of the outcomes was 5.

For each study, we estimated the aggregate data that might be used by an aggregate data meta-analysis: the sample mean, standard error of the mean, median, first and third quartiles, and minimum and maximum values. We will refer to the collection of 15 or 50 studies that might form a meta-analysis as a generated data set. We generated 1000 data sets for each combination of data generation parameters. Overall, there were $2^3 \times 13 = 104$ combinations of data generation parameters in the simulation study.

In our primary investigations, we considered that: (i) all studies present medians, first and third quartiles, and the number of subjects; (ii) all studies present medians, minimum and maximum values, and the number of subjects; or, (iii) all studies present sample means and their standard errors.

Each study was classified according to Bowley's coefficient of skewness[15] ($SK_b$), and we categorized these as low ($SK_b \leq 0.1$), medium ($0.1 < SK_b \leq 0.2$), high ($0.2 < SK_b \leq 0.4$), and very high ($SK_b > 0.4$). Bowley's coefficient of skewness is given by

$$SK_b = \frac{Q_1 - 2Q_2 + Q_3}{Q_3 - Q_1}$$

### 3.2. Sensitivity analysis

In sensitivity analyses, we considered that different studies might present different information in terms of the measure of interest (i.e., the mean or median) and in terms of the spread of that measure (i.e., the standard error of the mean, first and third quartiles, or minimum and maximum values). We sought to mimic real life more closely where a mix of means and medians is obtained from the source studies. Here, the data sets contained a mix of sample means with their standard errors and medians depending on whether the data were normally distributed according to the Shapiro-Wilk test[16]. When a median was reported, it was randomly selected whether the first and third quartiles or the minimum and maximum values were reported for the spread.

### 3.3. Estimating pooled measures

When all studies reported medians, we considered the median-based approaches (i.e., MM and WM) and the transformation-based approaches (i.e., T1 and T2) in the simulation study. For the median-based approaches, we estimated the pooled median and 95% CI as described in Section 2. For the transformation-based approaches, we estimated the sample mean and its standard error from each study and then estimated the fixed effect and random effects pooled mean and 95% CI, as described in Section 2.

When studies reported a mix of means and medians, we considered the same approaches as when all studies reported medians: the median-based approaches and the transformation-based approaches. For the median-based approaches, we treated the means as medians and then estimated the pooled median as described in Section 2. The transformation-based approaches estimated the sample mean and its standard error from each study that reported a median. Then, we estimated a fixed effect and random effects pooled mean using the study-specific sample means and their standard errors as well as the estimated sample means and their standard errors, as described in Section 2.

When all studies reported means, we considered the MEANS approaches as well as the median-based approaches. The median-based approaches treated the means as medians.

We note that when we treat the means as if they are medians for the median-based approaches, we assume that the means well approximate the medians (i.e., the outcome distributions is approximately symmetric in these studies). When this assumption does not hold, treating the means as medians will yield biased results. We address this point further in the Discussion.

### 3.4. Performance measures

We estimated the percent error, absolute percent error, and mean squared error of the pooled estimates. We defined the percent error as

$$percent\ error\ of\ x = \frac{estimated\ x - true\ x}{true\ x} \times 100$$

and the absolute percent error as the absolute value of the percent error. Although unbiased estimators will have percent error of zero, we note that unbiased estimators can have average absolute percent error that deviates considerably from zero. We estimated the coverage of the 95% CIs. We estimated the average $I^2$ and the average inter-study variance for the transformation-based approaches and the MEANS approaches.[17]

For the percent error, absolute percent error, mean squared error, and coverage, the true value of the pooled estimate depended on the approach used. In approaches estimating the true mean (i.e., T1, T2, and MEANS), we used the mean of the generating distribution as the true value. In the approaches estimating the true median (i.e., MM and WM), the median of the generating distribution was used as the truth.

### 3.5. Data analysis

We divided our analysis into three general scenarios: all studies reported medians, all studies reported means, and studies reported a mix of means and medians. To analyze the data in the simulation, we followed the method outlined by Chipman[18]. We first evaluated the extent to which the approach, mean skew, inter-study variance, number of studies, and median number of subjects per study affected the performance of the approaches in the three scenarios. This was measured by the overall median absolute percent error, percent error, mean squared error, and coverage calculated across each level of the factors. Then, we investigated how the approaches in these scenarios performed under each of the mean skew and inter-study variance levels. Finally, we analyzed the coverage of the 95% CIs of the approaches across each of the mean skew levels in each scenario.

4. Results

Initial analyses showed that number of studies and the median number of subjects per study did not considerably differentiate performance amongst the approaches, as measured by the absolute percent error (APE), percent error (PE), mean squared error (MSE), and coverage of the 95% CIs. Therefore, we restricted our analysis to the scenario where the number of studies was 50 and median number of subjects per study was 100. Similar results hold when fixing other values for the number of studies and median number of subjects per study. In Appendix A, B, and C, we present the median and first and third quartiles of the APE, PE, and MSE calculated across each combination of the mean skew and inter-study variance levels in the simulation. Additionally, the coverage of the 95% CIs calculated across each combination of the mean skew and inter-study variance levels is given in Appendix D.

*4.1. Percent error and absolute percent error*

When all studies reported medians, the median-based approaches overall performed better than the transformation-based approaches. The median-based approaches had considerably lower overall median APE for estimating the true median compared to the median APE using the transformation-based approaches for estimating the true mean. Furthermore, the median-based approaches had approximately 0% overall median PE. In the sensitivity analysis where a mix of means and medians were presented, the overall median APE and PE were very similar to the case where only medians were presented. However, when all studies reported sample means and their standard errors, treating the means as medians and using the median-based approaches estimated the true median poorly.

Figures 1 and 2 present APE or PE, respectively, of the approaches versus the mean skew when inter-study variance was fixed at $\tau^2 = 1/4$. With approximately symmetric data (mean $SK_b \leq 0.1$), the median-based approaches and random effects transformation-based approaches in the three scenarios yielded APE and PE near 0% for estimating the true median and mean, respectively. With highly skewed data (mean $SK_b > 0.40$), when only medians were presented, the median-based approaches outperformed the transformation-based approaches. The transformation-based approaches had median APE of 66% and higher for estimating the true mean, while all the median-based approaches had median APE less than 8% for estimating the true median. Furthermore, the median-based approaches had PE of approximately 0% in this scenario. When studies reported a mix of means and medians, the APE and PE values of the median-based approaches were nearly identical to when only medians were presented. Finally, when data were highly skewed (mean $SK_b > 0.40$) and each study presented a sample mean and its standard error, treating the means as medians and using the median-based methods resulted in high APE and PE for estimating the true median. In this scenario, the MEANS$_{RE}$ approach had median APE of 36% for estimating the true mean. Similar results were observed when fixing inter-study variance at $\tau^2 = 1/16$, $\tau^2 = 1$, or $\tau^2 = 4$.

Figures 3 and 4 show the APE or the PE, respectively, of the approaches versus the inter-study variance in all three scenarios when mean skew was high ($0.2 < \text{mean } SK_b \leq 0.4$). When all studies reported medians, the median-based approaches performed better than the transformation-based approaches in each of the inter-study variance levels, and performance was less affected by high inter-study variance compared to the transformation-based approaches. At the highest inter-study variance level ($\tau^2 = 4$), the median-based approaches had median APE under 30% for estimating the true median, but the transformation-based approaches all had median ABP greater than 80% for estimating the true mean. Moreover, the median-based approaches had approximately 0% median PE. The performance of these approaches did not considerably change when a mix of means and medians was reported. Lastly, we consider the case where all studies reported sample means and their standard errors. Treating the means as medians and using the median-based approaches resulted in high APE and PE for estimating the true median in each inter-study variance level. With low inter-study variance ($\tau^2 = 1/16$), the MEANS$_{RE}$ approach yielded a median APE of 5% for estimating the true mean. As inter-study variance increased, the APE increased and the PE deviated further from 0%. At the highest inter-study variance level in the simulation ($\tau^2 = 4$), the MEANS$_{RE}$ approach had median APE of 95% and median PE of -95%. Similar trends were observed when fixing the skewness at other levels.

### 4.2. Mean squared error

MSE did not differ considerably between the two median-based approaches. As expected, treating means as medians and using the median based approaches resulted in increased MSE. The overall median MSE for the MM and WM approaches was nearly 0 when all studies reported medians and was 1 when all studies reported means. For the transformation-based approaches, the T1$_{RE}$ had the lowest overall median MSE for estimating the true mean and the other transformation-based approaches had similar overall median MSE compared to each other. Results were similar when studies reported a mix of means and medians. When all studies reported means, the MEANS$_{RE}$ approach had lower overall MSE compared to the MEANS$_{FE}$ approach.

Figure C1 presents the MSE of the approaches versus the mean skew level when inter-study variance was fixed at $\tau^2 = 1/4$. In Figure C2, we display the MSE of the approaches versus the inter-study variance when we fixed the mean skew level to be high (0.2 < mean $SK_b$ ≤ 0.4). When all studies reported medians or when studies reported a mix of means and medians, the median-based methods had median MSE of nearly 0 for estimating the true median in all investigated scenarios in Figures C1 and C2. However, when all studies presented means along with either highly skewed data (mean $SK_b$ > 0.40) or data with high inter-study variance ($\tau^2 = 4$), treating the means as medians and using the median-based approaches resulted in high median MSE for estimating the true median. For the transformation-based methods, the random effects methods had lower median MSE compared to the fixed effect methods in most investigated scenarios. T2$_{RE}$ had very high median MSE when applied to highly skewed data (mean $SK_b$ > 0.40). The MSE of the transformation and median-based methods did not differ considerably whether only medians or a mix of mean and medians were presented. When all studies reported means, the MEANS$_{RE}$ approach had lower median MSE in each of the inter-study variance and mean skew levels compared to the MEANS$_{FE}$ approach. Similar results were observed for other combinations of data generation parameters.

*4.3. Coverage*

Figure 5 shows the coverage of the 95% CIs of the approaches across the four mean skew levels when we fixed inter-study variance at $\tau^2 = 1/4$. When all studies reported medians, the MM approach attained nominal or near nominal coverage, and the WM approach attained coverage between 88% and 89%, for the true median across all skew levels. For the random effects transformation-based approaches, the T1$_{RE}$ and T2$_{RE}$ approaches had coverage of 72% and 65% for the true mean when the mean skew was classified as low (mean $SK_b$ ≤ 0.1), and decreased towards 0% as the mean skew increased. The fixed effect transformation approaches had coverage of nearly 0% in all skew levels. When studies reported a mix of means and medians, the coverage of the median-based approaches was nearly identical to when all studies reported medians. The coverage increased slightly for the transformation-based approaches in this scenario compared to when all studies reported medians. When all studies reported means, with low skew (mean $SK_b$ ≤ 0.10), MEANS$_{RE}$ had coverage of 76% for the true mean, whereas MM and WM had coverage of 93% and 87%, respectively, for the true median. In the highest skew level (mean $SK_b$ > 0.4), all approaches had coverage of nearly 0%. The MEANS$_{FE}$ approach had approximately 0% coverage for the true mean in all skew levels. Similar results were observed at other levels of inter-study variance.

In the sensitivity analysis where studies reported either a mean or median based on the Shapiro-Wilk test, approximately 92% of the studies presented medians. The percentage of studies that reported medians in the low, medium, high, and very high skew levels were 61%, 87%, 99% and approximately 100%, respectively. As for inter-study variance, the percentage of studies that reported medians for $\tau^2 = 1/16$, $\tau^2 = 1/4$, $\tau^2 = 1$, and $\tau^2 = 4$ were 88%, 88%, 97% and nearly 100%, respectively.

*4.4. Worst-case performance*

We investigated the worst-case performance of the median-based methods when applied to meta-analyze studies that reported either only medians or a mix of means and medians. This occurred when the number of studies and number of subjects per study were at the lowest levels and when the inter-study variance and mean skew were at the highest levels. In this scenario, the method using the median of the study-specific medians yielded a median absolute percent error of 41%, median percent error of -0.2%, median MSE of 18, and coverage of 95%. The method using the weighted median of the study-specific medians resulted in a median absolute percent error of 44%, median percent error of -1.1%, median MSE of 21, and coverage of 94%. On the other hand, the MEANS$_{FE}$ approach, which represents the best-case scenario for a transformation-based approach in a fixed effect analysis, yielded a median absolute percent error of 99%, median percent error of -99%, median MSE of approximately 73,000, and coverage of 0%. The MEANS$_{RE}$ approach, the best-case scenario for a transformation-based approach in a random effects analysis, had median absolute percent error, median percent error, median MSE, and coverage of 97%, -97%, 69,000, and 0%, respectively.

5. **Example: Estimating Patient Delay in Pulmonary Tuberculosis Diagnosis**

We applied these methods to a data set collected to estimate various delays in pulmonary tuberculosis (TB) diagnosis and treatment in China.[2] We limited our analysis to the patient delay (PD), which is defined in Sreeramareddy et al[12] as the length of time between the onset of symptoms and the patient's first contact with a health care provider. The data set consisted of summary statistics from 50 studies, all of which reported the median PD. Forty-nine of the studies reported the number of subjects, and 34 included the minimum and maximum values, first and third quartiles, or both measures as the spread, which are presented in Figure 6. A large study consisting of over 40% of the total number of subjects in the meta-analysis presented an outlier median PD of 106.5 days (see Discussion). This study did not report a measure of spread.

For the median-based methods, we applied the MM method to all 50 studies that reported the median PD and we applied the WM method to the 49 studies that reported the number of subjects with the median PD. The analysis using the median-based methods was performed twice—one time including the outlier study and one time excluding this study. We present the results when the outlier study was excluded. The transformation-based methods were applied to the 34 studies that reported a measure of spread with the median PD. As the studies reported a mix of measures of spread, the transformation-based approaches were applied in the following way. If a study reported the minimum and maximum values as well as the first and third quartiles, we discarded the minimum and maximum values and only considered the first and third quartiles. Then, we applied the methods recommended by Wan et al[5] to estimate the sample means and its standard errors of each study. We used formulas (1) and (2) when studies reported first and third quartiles and formulas (3) and (4) when studies only reported minimum and maximum values as the spread, as demonstrated in Figure 6. We then estimated a fixed effect and random effects pooled mean, which we called TRANS$_{FE}$ and TRANS$_{RE}$, respectively. To evaluate the skewness of PD, we calculated Bowley's coefficient of skewness from all studies that reported the median and first and third quartiles. The mean Bowley's coefficient of skewness was 0.38, which indicates that the distribution of PD in these studies was highly skewed.

The MM method estimated a PD of approximately 16 days. The WM estimate was considerably larger (38 days) than the MM estimate because the two studies with the largest PDs had over half of the total number of subjects. When estimating the pooled mean with the TRANS$_{RE}$ method, there was considerable inter-study variance ($p < 0.0001$, $I^2 = 99.88\%$). The pooled mean estimates for the TRANS$_{FE}$ and TRANS$_{RE}$ approaches were 26 days and 105 days, respectively.

One reviewer suggested that we perform a subgroup analysis on the studies reporting the first and third quartiles. The simulation results indicate that the transformation-based method for studies reporting the minimum and maximum values (i.e., T2) is highly inaccurate and performs considerably worse than the transformation method for studies reporting the first and third quartiles (i.e., T1). Moreover, in Figure 6, we observe considerable variation in the estimated sample means for studies that only report the minimum and maximum values. For these reasons, we meta-analyze the 16 studies that report the first and third quartiles. In this case, the fixed effect and random effect transformation-based methods produce similar estimates for the patient delay: 21 days (95% CI: 21, 22) for the fixed effect approach and 25 days (95% CI: 15, 34) for the random effects approach. The MM and WM methods yielded estimates of 13 days (95% CI: 8, 20) and 60 days (95% CI: 20, 60), respectively. The large discrepancy between the MM and WM estimates was due to a study with a patient delay of 60 days and comprised approximately 70% of the total number of subjects in the subgroup.

## 6. Discussion

In this paper, we investigated approaches for meta-analyzing medians, considering the one-sample case. We demonstrated via simulation that whether meta-analyzing medians or a mix of means and medians, methods that use the study-specific medians performed better than transformation-based approaches, particularly when applied to skewed data or data with high inter-study variance. Pooled medians estimated via the median-based approaches had considerably lower absolute percent error than pooled means estimated via transformation-based approaches, while maintaining nominal or near nominal coverage, and percent error of approximately 0% across all investigated scenarios.

The median-based approaches offer several other advantages compared to the transformation-based approaches. These methods can be used on a wider range of studies, as they do not require the studies to report a measure of spread. Furthermore, they require no distributional assumptions, while the transformation-based methods of Wan et al[5] assume that the outcome variable is normally distributed. Moreover, the median-based methods are insensitive to correlations between the measure of effect and its variance, which has been shown by several authors[19-21] to cause substantial bias in meta-analyses using inverse variance weighting.

Several methods for estimating the sample mean and standard deviation (SD) from the median and various measures of spread have been proposed.[4-7] However, the median-based methods applied to data that consists of medians performed better than conducting a meta-analysis with the *actual* sample mean and standard error (SE) of the mean of each study. This represents the best-case scenario for approaches transforming medians into sample means and SDs. With approximately symmetric data with low inter-study variance, the median-based approaches performed comparably to using the actual sample means and their SEs. With highly skewed data or data with high inter-study variance, the median-based approaches had considerably lower median absolute percent error and median percent error closer to 0% than the approaches using the sample means and their SEs. The median-based approaches maintained coverage of about 90% and higher across all mean skew levels and inter-study variance levels, while the coverage of methods using the sample means and their SEs approached 0% in the highest mean skew and inter-study variance levels.

The median based approaches do not provide an estimate of inter-study heterogeneity. Estimates of between-study heterogeneity have other uses than for weighting in meta-analysis (e.g., indicating whether studies should be summarized together at all[22]). However, the transformation-based approaches estimated heterogeneity poorly in the simulation study (data not shown), which is likely to be common when the underlying outcomes are not normally distributed.

The scenario where a mix of means and medians was reported based on the results of a normality test was meant to mirror real life where researchers may report a mean or a median based on the skewness of the data. Because the reported means were good approximations of medians, the performance of the median-based methods in this scenario was nearly identical to the case where all studies reported medians. If a sample mean is reported even if the data are skewed we would expect the performance of the median-based approaches to be somewhat worse than in the simulated mixed scenario because the reported means may no longer be good approximations of medians.

There are some limitations to the work presented here. We generated the outcome variable to mimic time-based outcomes, which often follow log-normal distributions.

However, others[4-7] considered additional distributions when investigating the estimation of the sample mean and SD from the median and spread. Kwon and Reis[7] compared the performance of approaches to estimate the sample mean and SD under various distributions and concluded that the distribution of the outcome variable, especially skewness, strongly influenced performance. Therefore, considering just the log-normal distribution in our simulation limits the generalizability of our results, as the performance of the approaches may change when applied to data generated from other distributions. However, we observed the performance of the approaches when applied to data varying from approximately symmetric to substantially skewed by varying the parameters of the log normal distribution. Consequently, we expect that the performance of the median-based approaches and transformation-based approaches would decrease as the skewness of the distribution increases, and that the performance of the transformation-based approaches would be more strongly affected, as observed in the simulations. Moreover, when meta-analyzing studies that report the median in real life, the underlying distribution of the outcome is typically unknown. Thus, we used methods for estimating the sample mean and SD that are not specific to the log-normal context.

Although there are several methods to estimate the sample mean and SD from the median and reported spread,[4-7] we considered the methods recommended by Wan et al[5]. Kwon and Reis[7] showed that these methods performed the best overall when data were generated from distributions and sample sizes comparable to those used in our simulation study, and estimated the SD the best. Although the Approximate Bayesian Computation method of Kwon and Reis[7] performed better than that those recommended by Wan et al[5] for estimating the sample mean, it is very computationally intensive and requires knowledge of the underlying distribution that generates the data *a priori*. Some would consider the method of Bland[6] to be the definitive approach, but Bland's method[6] requires studies to report all of the following summary measures: sample size, median, first and third quartiles, and minimum and maximum values. Only three out of the 50 studies in our example data set reported all of these. As such, we did not include Bland's method[6].

Based on the results of the simulation study, we strongly encourage authors, at the very least, to report a mean or median of the outcome variable in their studies based on the skewness of the data. Preferably, authors should report means *and* medians in their studies, as the accuracy of the pooled estimates of meta-analyses applied on these studies considerably increases with more information. In most scenarios in our simulation, we observed that the median-based approaches applied to data with a mix of means and medians based on the skewness of the data performed considerably better than a standard random effects meta-analysis using the sample mean and SE of each study.

When data analysts are faced with incorporating medians in meta-analyses, they may consider using the transformation-based approaches or the median-based approaches. As demonstrated via simulation, the relative performance of the transformation-based approaches and the median-based approaches depended heavily on the skewness of the data. Therefore, we recommend that data analysts select the approach to meta-analyze the data based on the skewness. Bowley's coefficient of skewness ($SK_b$) can be calculated from studies that report the median and first and third quartiles, which are commonly reported measures of spread for the median. Based on the results of the simulation study, we suggest using the median-based approaches if the mean $SK_b$ is greater than 0.1 and expect all approaches to perform well and similarly for lower values. If individual patient data is available from studies in the meta-analysis, performing a transformed analysis on the normal scale is possible, as are several other approaches.[23,24]

Our simulation results suggest that the median-based approaches are most suitable for estimating the patient delay in tuberculosis diagnosis in China. This is because all studies reported medians, data were highly skewed (mean $SK_b = 0.38$), and there was important inter-study variability calculated from the estimated means and SEs. The analysis using the median-based methods was performed twice due to an outlier study with a large number of subjects and high patient delay. The presence of the outlier study did not considerably affect the pooled estimate of the unweighted median method but resulted in a considerable difference for the pooled estimate of the weighted median method. When including the outlier study, we estimated a patient delay of 18 days (95% CI: 10, 21) by taking the median of the study-specific medians and 60 days (95% CI: 38, 107) using the weighted median of the study-specific medians. When excluding the outlier study, the unweighted and weighted medians of the study-specific medians were 16 days (95% CI: 10, 21) days and 38 days (95% CI: 38, 38), respectively. In both cases, the median-based approaches produced vastly different pooled estimates of the median patient delay. We investigated instances when there was a discrepancy between the unweighted and weighted median methods in the simulation study. When the weighted median was greater than twice the median, the median method had considerably lower absolute percent error than the weighted median method. Moreover, where there was high inter-study variance in the simulation study, the median method performed better than the weighted median method, as high inter-study variance suggests that different studies should have approximately equal weights. Lastly, when all studies reported medians and data were highly skewed in the simulation study, the median method attained higher coverage and lower absolute percent error and MSE than the weighted median method. We conclude with an estimate of 16 days (95% CI: 10, 21) for patient delay in pulmonary tuberculosis diagnosis in China.

In conclusion, we compared the performance of two median-based approaches and two transformation-based approaches to meta-analyze data that consists of medians. We evaluated their performance over a wide range of scenarios meant to mirror real-life. Moreover, we examined how these approaches performed whether all studies reported medians, studies reported a mix of means and medians, or all studies reported means. When meta-analyzing data that consists of medians, the median-based approaches performed better than the transformation-based approaches in nearly all investigated scenarios and better than the best-case scenario for transformation approaches when applied to heavily skewed data or data with high inter-study variance. We encourage authors and data analysts to follow the guidelines above. In future work, we investigate methods to estimate the sampling variance of the median from commonly reported summary measures. This will allow us to pool medians using the inverse variance method frequently used in meta-analysis.

**Acknowledgements**

Andrea Benedetti is supported by the FRQS.

# References

1. Bastian H, Glasziou P, Chalmers I. Seventy-five trials and eleven systematic reviews a day: how will we ever keep up? *PLoS Med.* 2010;7(9):e1000326.
2. Qin ZZ. *Delays in diagnosis and treatment of pulmonary tuberculosis, and patient care-seeking pathways in China: a systematic review and meta-analysis* [master's thesis]. Montreal, Canada, McGill University; 2015.
3. Sohn H. *Improving tuberculosis diagnosis in vulnerable populations: impact and cost-effectiveness of novel, rapid molecular assays* [dissertation]. Montreal, Canada, McGill University; 2016.
4. Hozo SP, Djulbegovic B, Hozo I. Estimating the mean and variance from the median, range, and the size of a sample. *BMC Med Res Methodol.* 2005;5:13.
5. Wan X, Wang W, Liu J, Tong T. Estimating the sample mean and standard deviation from the sample size, median, range and/or interquartile range. *BMC Med Res Methodol.* 2014;14:135.
6. Bland M. Estimating mean and standard deviation from the sample size, three quartiles, minimum, and maximum. *International Journal of Statistics in Medical Research.* 2015;4:57-64.
7. Kwon D, Reis IM. Simulation-based estimation of mean and standard deviation for meta-analysis via Approximate Bayesian Computation (ABC). *BMC Med Res Methodol.* 2015;15:61.
8. DerSimonian R, Kacker R. Random-effects model for meta-analysis of clinical trials: An update. *Contemporary Clinical Trials.* 2007;28(2):105-114.
9. Viechtbauer W. Bias and efficiency of meta-analytic variance estimators in the random-effects model. *J Educ Behav Stat.* 2005;30(3):261-293.
10. Hedges LV, Vevea JL. Fixed- and random-effects models in meta-analysis. *Psychological Methods.* 1998;3:486-504.
11. DerSimonian R, Laird N. Meta-analysis in clinical trials. *Control Clin Trials.* 1986;7(3):177-188.
12. Sreeramareddy CT, Panduru KV, Menten J, Van den Ende J. Time delays in diagnosis of pulmonary tuberculosis: a systematic review of literature. *BMC Infect Dis.* 2009;9:91.
13. Sreeramareddy CT, Qin ZZ, Satyanarayana S, Subbaraman R, Pai M. Delays in diagnosis and treatment of pulmonary tuberculosis in India: a systematic review. *Int J Tuberc Lung Dis.* 2014;18(3):255-266.
14. Conover WJ. *Practical Nonparametric Statistics.* 2nd ed. New York: John Wiley & Sons; 1980.
15. Kenney JF, Keeping ES. *Mathematics of Satistics, Part 1.* 3rd ed. Princeton, NJ: Van Nostrand; 1962.
16. Shapiro SS, Wilk MB. An analysis of variance test for normality (complete samples). *Biometrika.* 1965;52:591-611.
17. Higgins JP, Thompson SG. Quantifying heterogeneity in a meta-analysis. *Stat Med.* 2002;21(11):1539-1558.
18. Chipman H. Simulation studies for statistical procedures: Why can't we practice what we preach? Talk presented at: 43rd Annual Meeting of the Statistical Society of Canada; 2015; Halifax, Canada.
19. Shuster JJ. Empirical vs natural weighting in random effects meta-analysis. *Stat Med.* 2010;29(12):1259-1265.

20. Emerson JD HD, Mosteller. A comparison of procedures for combining risk differences in sets of 2×2 tables from clinical trials. *Journal of the Italian Statistical Society.* 1993;2:269-290.
21. Emerson JD, Hoaglin DC, Mosteller F. Simple robust procedures for combining risk differences in sets of 2 x 2 tables. *Stat Med.* 1996;15(14):1465-1488.
22. Higgins JP, Thompson SG, Deeks JJ, Altman DG. Measuring inconsistency in meta-analyses. *BMJ.* 2003;327(7414):557-560.
23. O'Rourke K. *The combining of information: Investigating and synthesizing what is possibly common in clinical observations or studies via likelihood* [dissertation]. Oxford, England, University of Oxford; 2008.
24. Geraci M, Bottai M. Linear quantile mixed models. *Statistics and Computing.* 2014;24(3):461-479.

**Figure 1.** Interaction plots of the primary and sensitivity analyses of the approach by the mean skew level. Performance is measured by APE.

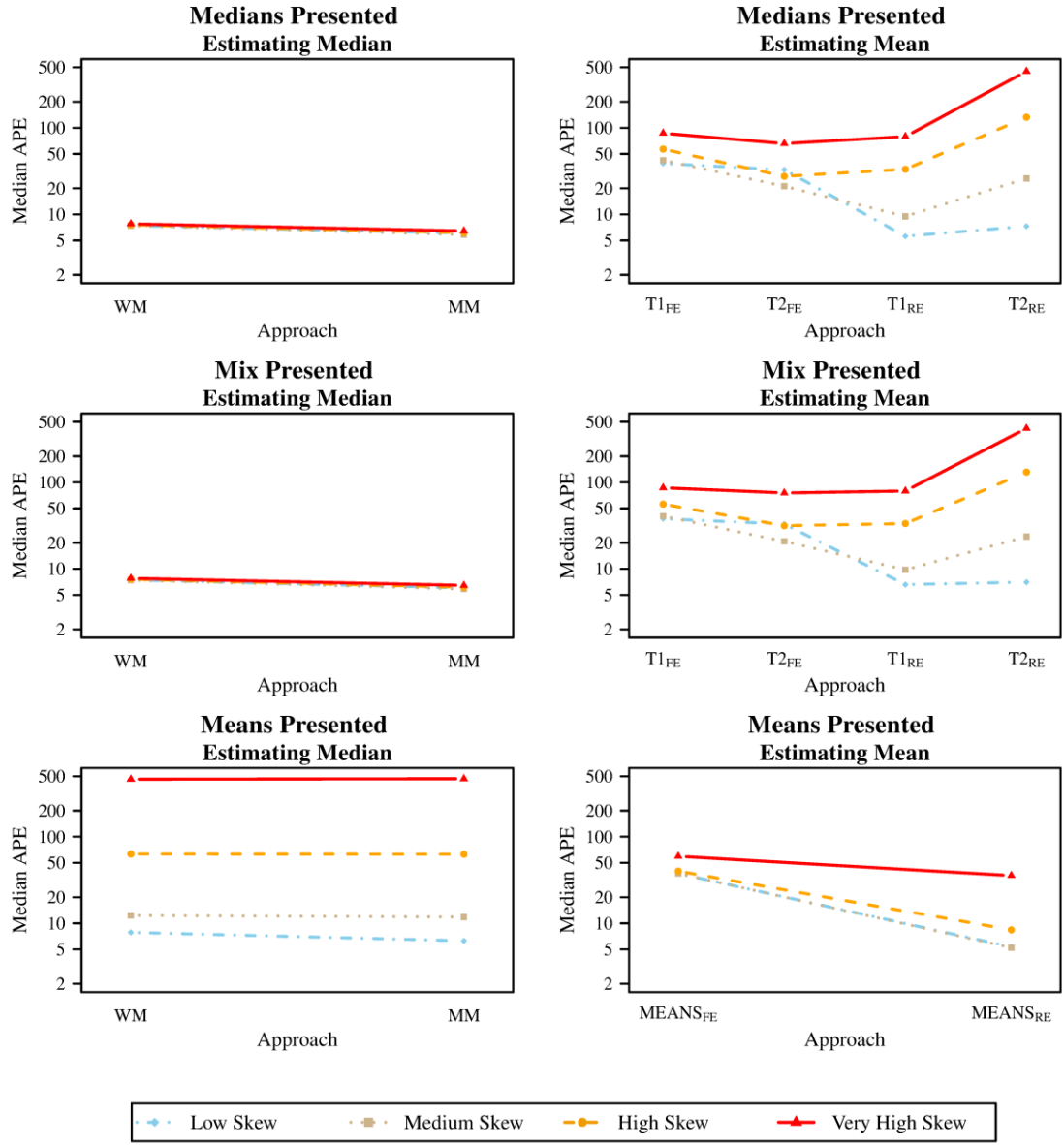

The APE was calculated over the 1,000 generated data sets for each combination of data generation parameters. The data set was restricted to the scenario where number of studies equalled 50, the median number of subjects per study equalled 100, and inter-study variance equalled 1/4. Studies with APE greater than 500% were removed. The y-coordinates are plotted on a log scale in all plots in this figure.

**Figure 2.** Interaction plots of the primary and sensitivity analyses of the approach by the mean skew level. Performance is measured by PE.

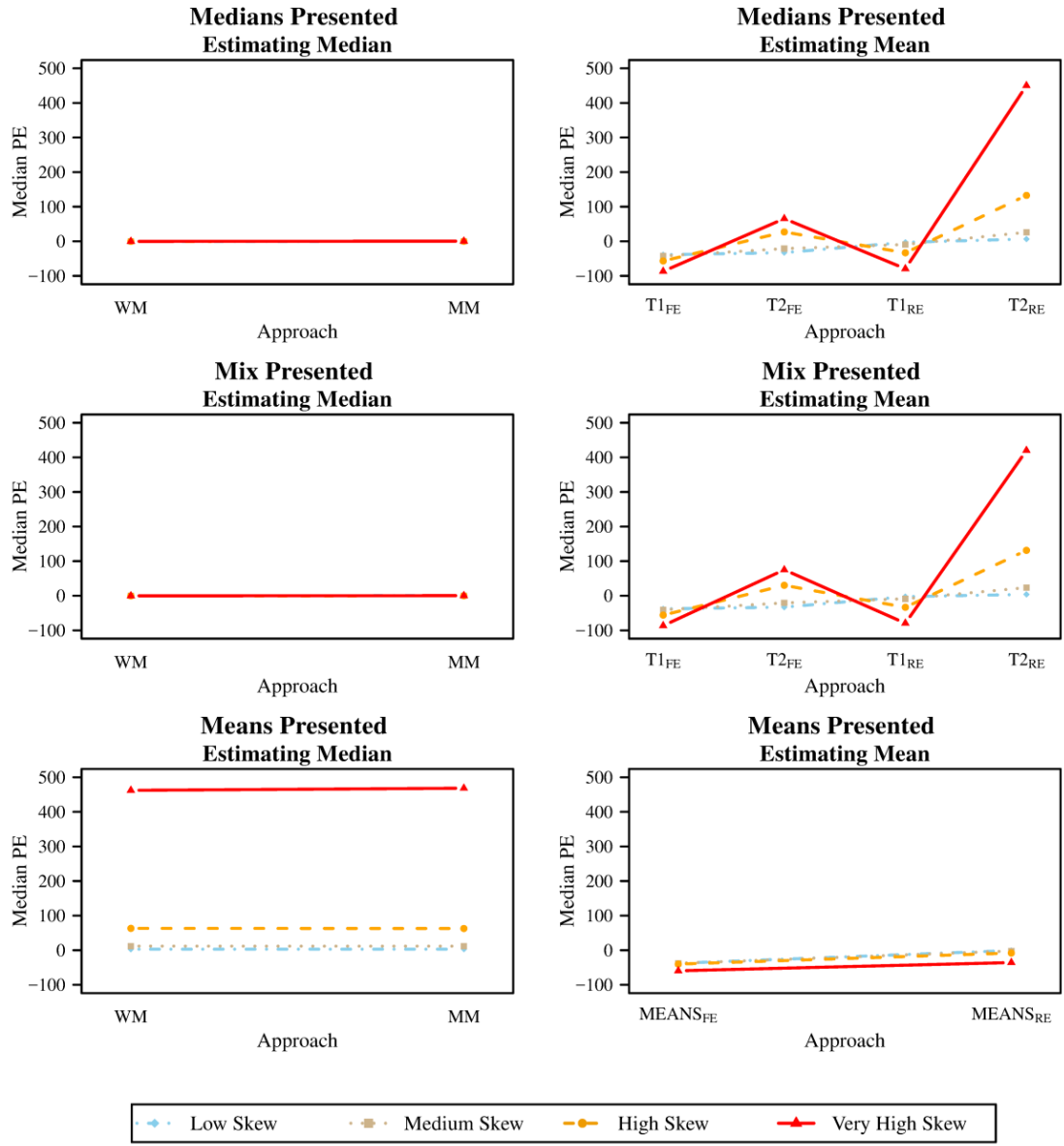

The PE was calculated over the 1,000 generated data sets for each combination of data generation parameters. The data set was restricted to the scenario where number of studies equalled 50, the median number of subjects per study equalled 100, and inter-study variance equalled 1/4. Studies with APE greater than 500% were removed.

**Figure 3.** Interaction plots of the primary and sensitivity analyses of the approach by the inter-study variance. Performance is measured by APE.

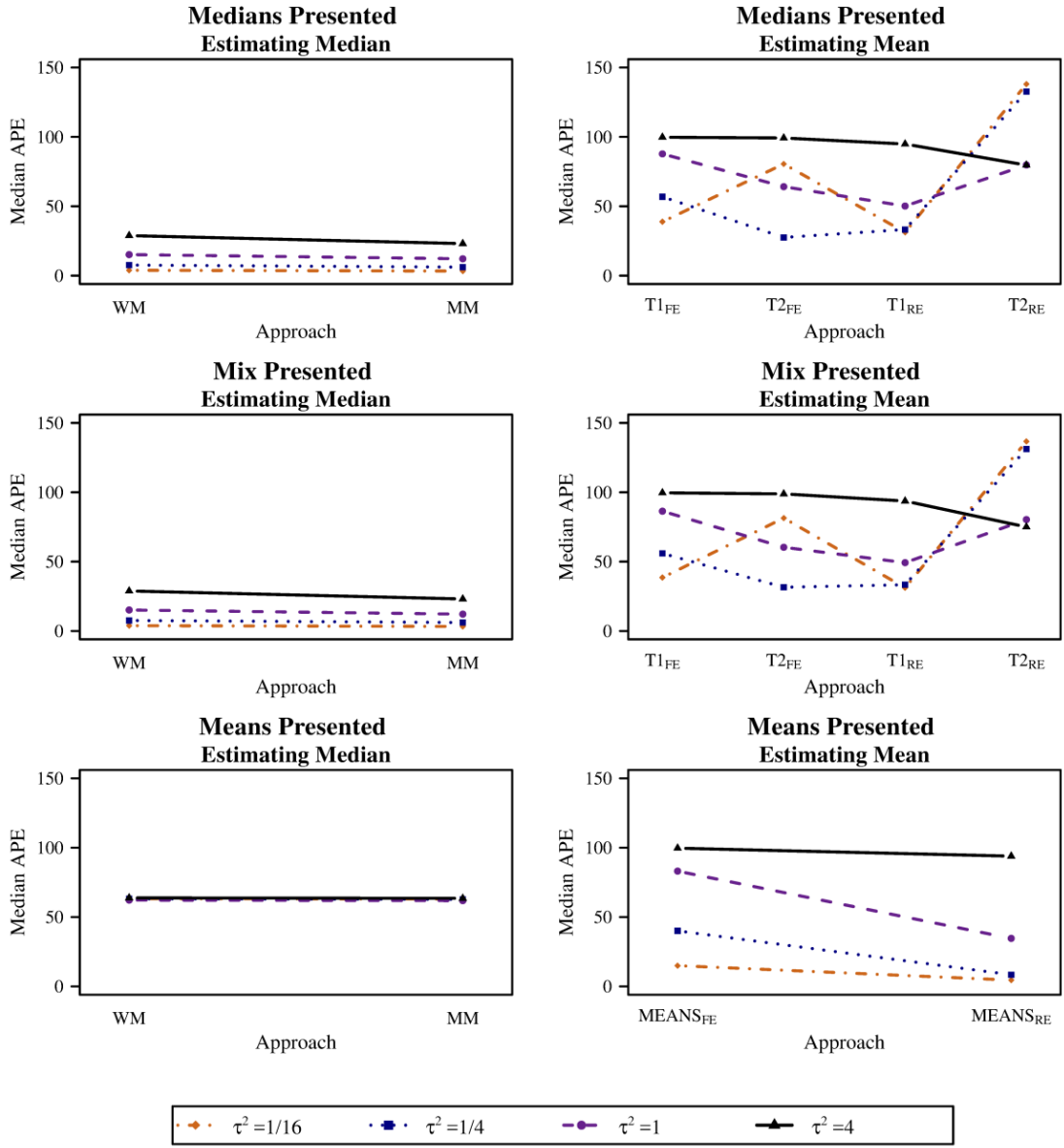

The APE was calculated over the 1,000 generated data sets for each combination of data generation parameters. The data set was restricted to the scenario where number of studies equalled 50, the median number of subjects per study equalled 100, and the mean skew level was classified as high ($0.2 <$ mean $SK_b \leq 0.4$). Studies with APE greater than 500% were removed.

**Figure 4.** Interaction plots of the primary and sensitivity analyses of the approach by the inter-study variance. Performance is measured by PE.

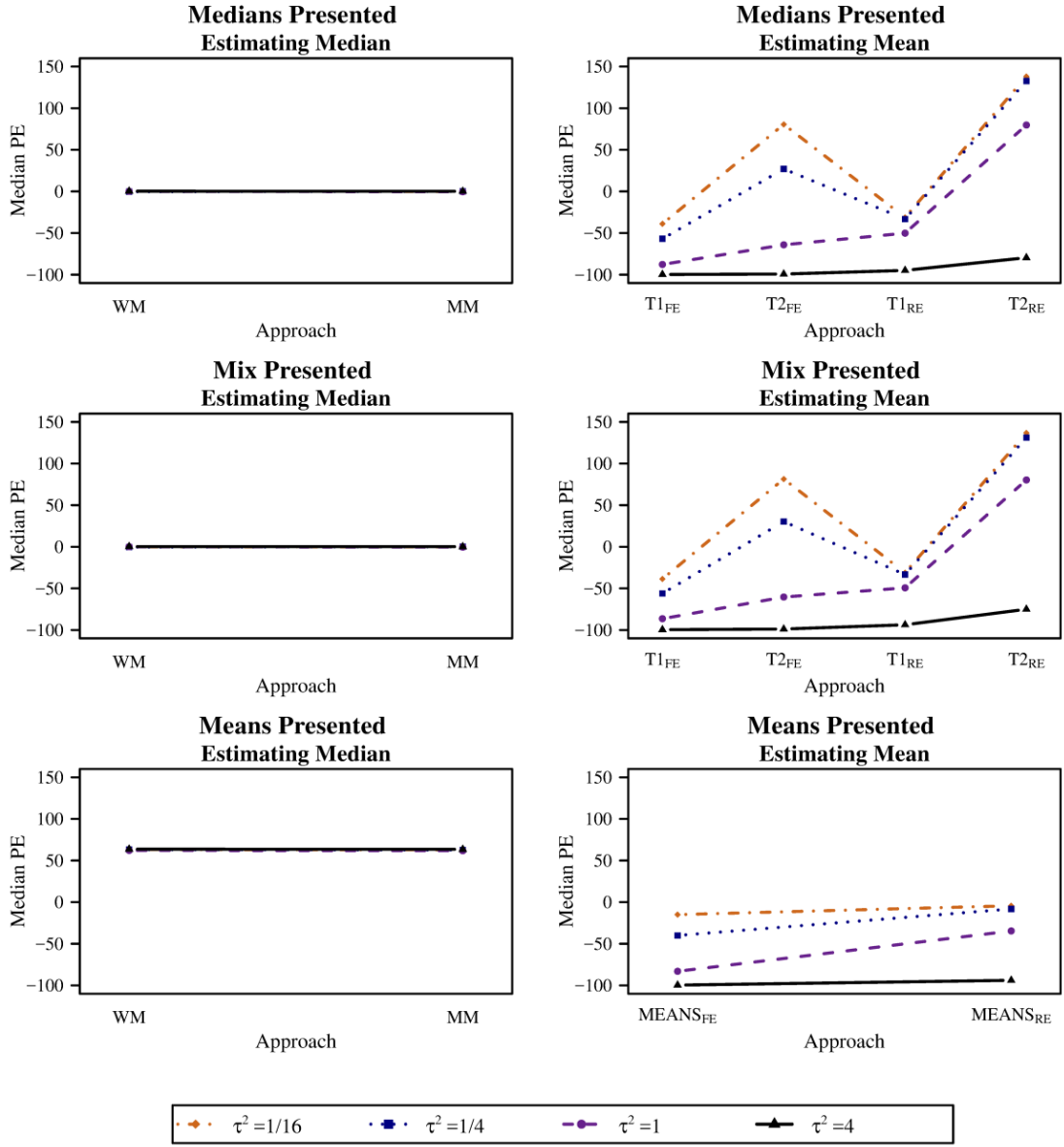

The PE was calculated over the 1,000 generated data sets for each combination of data generation parameters. The data set was restricted to the scenario where number of studies equalled 50, the median number of subjects per study equalled 100, and the mean skew level was classified as high ($0.2 < $ mean $SK_b \leq 0.4$). Studies with APE greater than 500% were removed.

**Figure 5.** Coverage probabilities of the approaches by the mean skew.

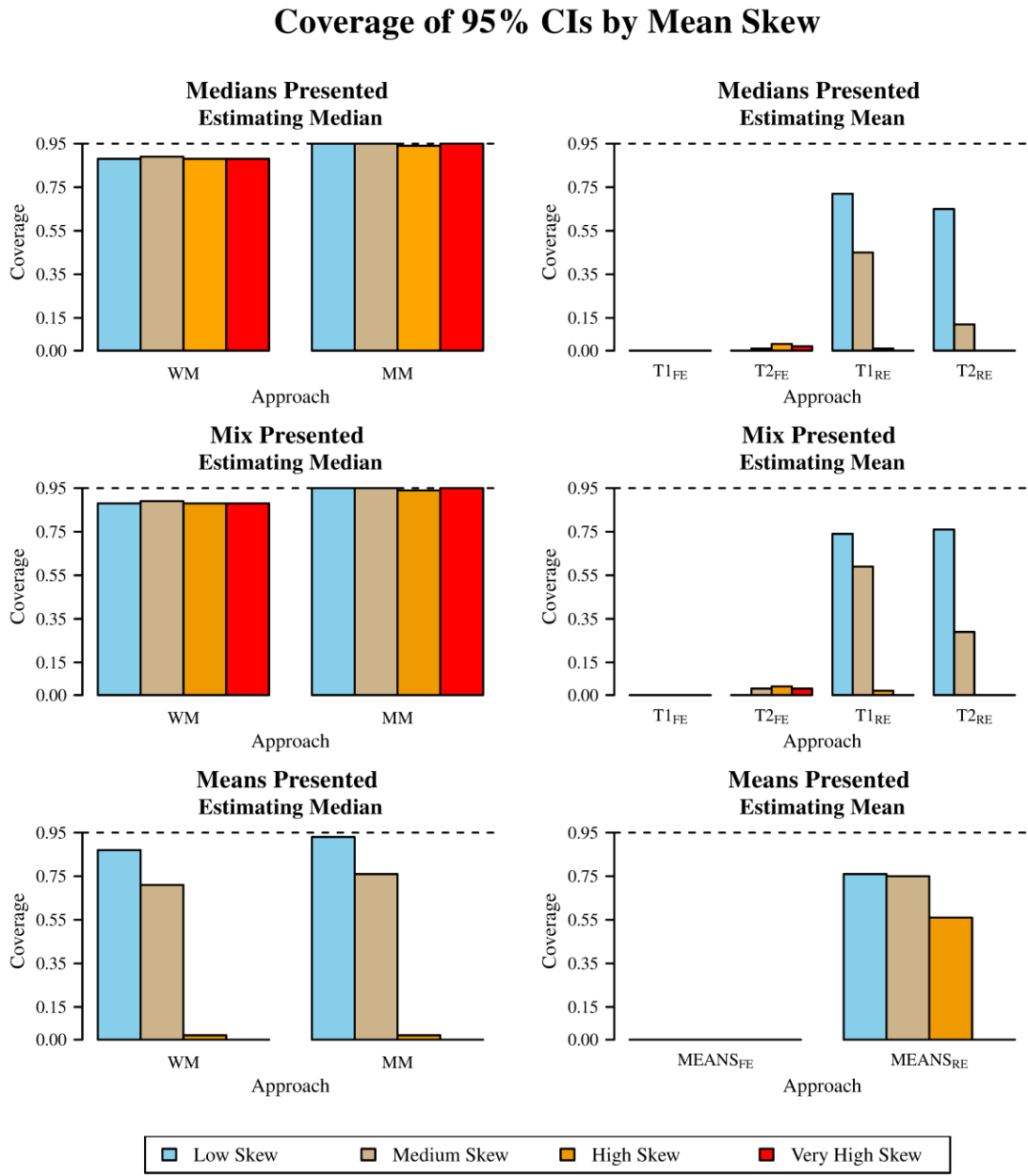

The coverage was calculated over the 1,000 generated data sets for each combination of data generation parameters. The data set was restricted to the scenario where number of studies was 50, the median number of subjects per study was 100, and the inter-study variance equalled 1/4. Studies with absolute percent error greater than 500% were removed.

**Figure 6.** Study-specific summary statistics of patient delay (PD) in the meta-analysis and their transformation to a sample mean and standard deviation via the methods recommended by Wan et al[5]. The 34 studies that reported a measure of spread along with the median PD are displayed. Blue markers were used when medians were reported in the study with first and third quartiles. For these studies, the ends of the error bars display the first and third quartiles of PD reported. Red markers were used when only the minimum and maximum values were reported as the spread. The ends of the error bars display the minimum and maximum values of PD reported for these studies. Pooled estimates of PD by median-based approaches and their 95% CIs are on the left and pooled estimates of PD by transformation-based approaches and their 95% CIs are on the right.

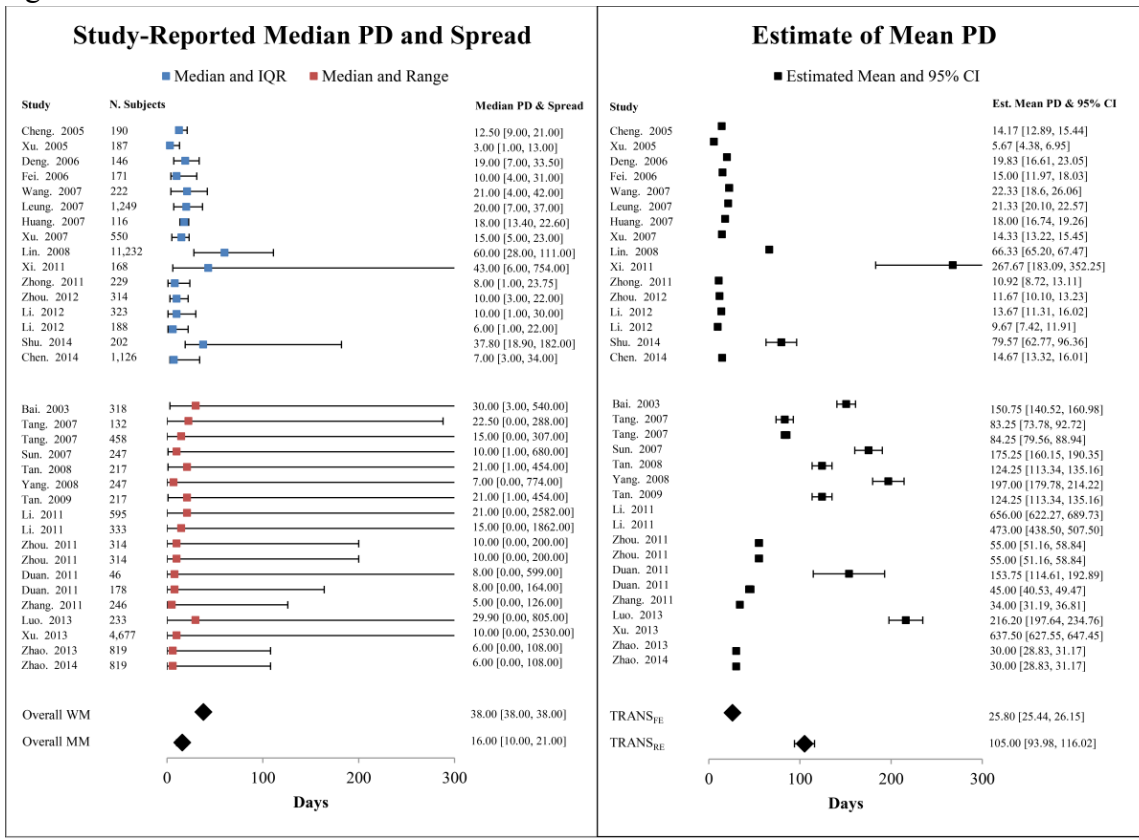

## Appendix A

The tables below present the median and first and third quartiles of the absolute percent error calculated across each combination of inter-study variance and mean skew levels in the primary and sensitivity analyses. The data set was restricted to the scenario where the number of studies was 50 and the median number of subjects per study was 100. Studies with absolute percent error greater than 500% were removed. The format for the entries of the following table is: median (first quartile, third quartile). Since the skewness was measured from the simulated data, not all combinations of mean skew levels and inter-study variance levels were observed. Entries with NA correspond to scenarios that were not observed in the simulation study.

**Table A1.** Median-based approaches.

| Approach | $\tau^2$ | Skew | Medians Given | Mix Given | Means Given |
|---|---|---|---|---|---|
| MM | 1/16 | Low | 2.92 (1.44, 5.12) | 3.02 (1.43, 5.16) | 3.81 (1.85, 6.40) |
| | | Medium | 3.00 (1.43, 5.20) | 3.01 (1.39, 5.15) | 11.96 (7.82, 15.81) |
| | | High | 3.28 (1.63, 5.67) | 3.28 (1.64, 5.69) | 63.00 (57.19, 68.57) |
| | | Very High | 4.12 (1.89, 7.03) | 4.12 (1.89, 7.03) | 476.16 (459.40, 489.17) |
| | 1/4 | Low | 6.07 (2.89, 10.12) | 6.03 (2.91, 10.10) | 6.28 (2.95, 11.09) |
| | | Medium | 5.84 (2.79, 9.94) | 5.86 (2.80, 9.97) | 11.82 (5.93, 18.68) |
| | | High | 6.14 (2.96, 10.64) | 6.13 (2.97, 10.64) | 62.73 (52.81, 73.67) |
| | | Very High | 6.44 (3.05, 11.24) | 6.44 (3.05, 11.24) | 468.29 (440.65, 485.45) |
| | 1 | Low | 7.46 (2.79, 15.78) | 7.46 (2.79, 15.78) | 19.39 (11.30, 22.87) |
| | | Medium | 12.02 (5.61, 20.09) | 12.02 (5.52, 20.12) | 15.77 (7.30, 27.81) |
| | | High | 12.14 (5.92, 20.15) | 12.16 (5.93, 20.14) | 61.78 (43.25, 83.28) |
| | | Very High | 12.43 (5.67, 20.78) | 12.43 (5.67, 20.78) | 441.53 (398.60, 471.30) |
| | 4 | Low | NA | NA | NA |
| | | Medium | NA | NA | NA |
| | | High | 23.10 (11.17, 39.80) | 23.09 (11.16, 39.80) | 63.47 (31.10, 107.97) |
| | | Very High | 24.20 (11.68, 40.25) | 24.20 (11.68, 40.25) | 384.45 (316.45, 443.05) |
| WM | 1/16 | Low | 3.84 (1.75, 6.39) | 3.85 (1.77, 6.44) | 4.39 (2.15, 7.57) |
| | | Medium | 3.83 (1.78, 6.58) | 3.84 (1.80, 6.60) | 11.88 (7.19, 16.81) |
| | | High | 3.85 (1.85, 6.67) | 3.84 (1.85, 6.67) | 63.26 (57.24, 70.20) |
| | | Very High | 4.64 (2.07, 7.98) | 4.64 (2.07, 7.98) | 479.15 (466.13, 490.90) |
| | 1/4 | Low | 7.44 (3.62, 12.73) | 7.45 (3.62, 12.81) | 7.85 (3.59, 13.31) |
| | | Medium | 7.40 (3.53, 12.62) | 7.40 (3.53, 12.63) | 12.32 (5.79, 20.80) |
| | | High | 7.59 (3.61, 13.12) | 7.59 (3.59, 13.12) | 63.14 (50.69, 76.74) |
| | | Very High | 7.78 (3.51, 13.58) | 7.78 (3.51, 13.58) | 462.21 (436.70, 482.55) |
| | 1 | Low | 16.99 (2.21, 55.08) | 16.99 (2.21, 55.08) | 34.06 (15.45, 77.13) |
| | | Medium | 14.76 (7.25, 25.38) | 14.76 (7.25, 25.39) | 17.89 (8.12, 32.37) |
| | | High | 15.16 (7.12, 25.25) | 15.16 (7.13, 25.25) | 62.17 (39.41, 89.64) |

|   | | | | |
|---|---|---|---|---|
|   | Very High | 15.73 (7.50, 26.50) | 15.73 (7.50, 26.50) | 433.11 (385.05, 469.17) |
| 4 | Low | NA | NA | NA |
|   | Medium | NA | NA | NA |
|   | High | 28.86 (13.63, 49.47) | 28.86 (13.63, 49.47) | 63.79 (28.11, 117.78) |
|   | Very High | 29.14 (14.65, 48.79) | 29.14 (14.65, 48.79) | 362.54 (279.24, 431.46) |

**Table A2.** Transformation-based approaches. In the column "Means Given", we present the absolute percent error of the MEANS$_{FE}$ approach for the fixed effect transformation-based methods and the absolute percent error of the MEANS$_{RE}$ approach for the random effects transformation-based methods.

| Approach | $\tau^2$ | Skew | Medians Given | Mix Given | Means Given |
|---|---|---|---|---|---|
| T1$_{FE}$ | 1/16 | Low | 13.19 (10.42, 16.08) | 12.73 (8.90, 16.17) | 11.33 (8.50, 14.27) |
| | | Medium | 18.44 (15.35, 21.54) | 17.91 (13.96, 21.78) | 11.85 (8.94, 14.88) |
| | | High | 38.87 (36.66, 41.10) | 38.56 (35.14, 41.55) | 14.95 (11.95, 18.02) |
| | | Very High | 81.19 (80.38, 81.98) | 81.07 (79.95, 82.19) | 43.07 (38.88, 47.65) |
| | 1/4 | Low | 38.81 (33.80, 44.17) | 37.90 (31.57, 44.35) | 37.60 (32.42, 42.84) |
| | | Medium | 42.24 (37.26, 47.51) | 40.73 (34.45, 47.29) | 37.68 (32.48, 43.23) |
| | | High | 56.83 (53.22, 60.64) | 55.99 (50.70, 60.98) | 40.02 (35.00, 45.39) |
| | | Very High | 86.73 (85.58, 87.94) | 86.48 (84.81, 88.14) | 59.45 (54.33, 64.64) |
| | 1 | Low | 81.42 (79.37, 82.61) | 77.31 (74.31, 82.71) | 79.90 (76.41, 81.15) |
| | | Medium | 83.74 (79.62, 87.81) | 82.10 (76.38, 87.08) | 82.33 (77.91, 86.80) |
| | | High | 87.77 (84.59, 90.78) | 86.38 (82.04, 90.29) | 83.08 (78.64, 87.18) |
| | | Very High | 96.21 (95.25, 97.16) | 95.75 (94.35, 96.96) | 87.92 (84.25, 91.23) |
| | 4 | Low | NA | NA | NA |
| | | Medium | NA | NA | NA |
| | | High | 99.75 (99.53, 99.88) | 99.61 (99.21, 99.83) | 99.64 (99.34, 99.83) |
| | | Very High | 99.92 (99.85, 99.96) | 99.88 (99.76, 99.95) | 99.73 (99.49, 99.87) |
| T2$_{FE}$ | 1/16 | Low | 4.80 (2.46, 7.91) | 6.03 (2.87, 9.72) | 11.33 (8.50, 14.27) |
| | | Medium | 13.41 (8.38, 18.11) | 12.90 (7.15, 19.08) | 11.85 (8.94, 14.88) |
| | | High | 80.50 (70.03, 90.93) | 81.44 (66.18, 96.62) | 14.95 (11.95, 18.02) |
| | | Very High | 127.92 (92.23, 163.43) | 134.48 (87.77, 189.80) | 43.07 (38.88, 47.65) |
| | 1/4 | Low | 32.76 (27.20, 38.41) | 32.96 (26.11, 39.61) | 37.60 (32.42, 42.84) |
| | | Medium | 21.16 (13.53, 29.14) | 20.80 (11.79, 30.07) | 37.68 (32.48, 43.23) |
| | | High | 27.54 (15.59, 40.49) | 31.53 (16.36, 48.46) | 40.02 (35.00, 45.39) |
| | | Very High | 65.80 (34.07, 101.10) | 75.45 (36.43, 124.27) | 59.45 (54.33, 64.64) |
| | 1 | Low | 72.06 (69.25, 76.50) | 71.11 (70.58, 72.99) | 79.90 (76.41, 81.15) |
| | | Medium | 76.98 (71.24, 82.76) | 74.86 (66.72, 81.82) | 82.33 (77.91, 86.80) |
| | | High | 64.13 (53.85, 73.28) | 60.37 (45.74, 71.92) | 83.08 (78.64, 87.18) |
| | | Very High | 48.61 (30.00, 65.25) | 44.77 (23.55, 64.92) | 87.92 (84.25, 91.23) |
| | 4 | Low | NA | NA | NA |
| | | Medium | NA | NA | NA |
| | | High | 99.23 (98.57, 99.64) | 98.83 (97.69, 99.48) | 99.64 (99.34, 99.83) |
| | | Very High | 98.73 (97.48, 99.43) | 97.96 (95.61, 99.14) | 99.73 (99.49, 99.87) |
| T1$_{RE}$ | 1/16 | Low | 2.95 (1.40, 4.89) | 3.33 (1.59, 5.48) | 2.42 (1.15, 4.00) |
| | | Medium | 8.44 (5.63, 10.96) | 7.96 (4.56, 11.32) | 2.60 (1.26, 4.40) |
| | | High | 31.24 (29.16, 33.01) | 31.19 (28.26, 33.78) | 4.51 (2.27, 6.94) |
|  |  | Very High | 78.16 (77.43, 78.85) | 78.12 (77.07, 79.16) | 28.79 (25.52, 31.87) |
|  | 1/4 | Low | 5.59 (2.62, 9.21) | 6.58 (3.18, 11.16) | 5.33 (2.40, 8.78) |
|  |  | Medium | 9.49 (4.97, 14.02) | 9.77 (4.93, 15.55) | 5.22 (2.51, 8.78) |
|  |  | High | 33.24 (29.61, 36.73) | 33.37 (28.13, 38.17) | 8.40 (4.36, 13.07) |
|  |  | Very High | 79.34 (78.20, 80.50) | 79.40 (77.71, 80.95) | 35.61 (30.80, 39.64) |
|  | 1 | Low | 14.40 (10.46, 18.10) | 14.02 (7.04, 20.23) | 8.82 (4.05, 12.57) |
|  |  | Medium | 25.45 (17.08, 32.42) | 25.65 (15.23, 35.04) | 19.95 (11.67, 27.24) |
|  |  | High | 50.15 (44.62, 55.11) | 49.27 (40.96, 56.37) | 34.60 (27.42, 41.39) |
|  |  | Very High | 86.24 (84.60, 87.71) | 85.78 (83.47, 87.93) | 62.77 (57.81, 67.39) |
|  | 4 | Low | NA | NA | NA |
|  |  | Medium | NA | NA | NA |
|  |  | High | 94.87 (93.19, 96.18) | 93.69 (91.03, 95.70) | 93.89 (91.82, 95.49) |
|  |  | Very High | 98.82 (98.40, 99.14) | 98.54 (97.82, 99.02) | 97.36 (96.32, 98.14) |
| T2$_{RE}$ | 1/16 | Low | 7.05 (4.50, 9.71) | 4.90 (2.35, 8.00) | 2.42 (1.15, 4.00) |
|  |  | Medium | 28.54 (23.12, 32.64) | 26.20 (18.91, 32.46) | 2.60 (1.26, 4.40) |
|  |  | High | 138.05 (125.67, 150.75) | 136.66 (119.41, 156.28) | 4.51 (2.27, 6.94) |
|  |  | Very High | 466.09 (433.28, 483.46) | 439.24 (391.85, 472.36) | 28.79 (25.52, 31.87) |
|  | 1/4 | Low | 7.31 (3.37, 12.44) | 7.03 (3.18, 12.14) | 5.33 (2.40, 8.78) |
|  |  | Medium | 26.02 (17.42, 33.62) | 23.64 (13.35, 33.78) | 5.22 (2.51, 8.78) |
|  |  | High | 132.70 (115.38, 149.19) | 131.23 (108.51, 156.26) | 8.40 (4.36, 13.07) |
|  |  | Very High | 450.02 (411.75, 478.19) | 420.10 (365.88, 461.99) | 35.61 (30.80, 39.64) |
|  | 1 | Low | 23.13 (19.46, 27.04) | 17.39 (15.29, 27.76) | 8.82 (4.05, 12.57) |
|  |  | Medium | 12.13 (5.58, 21.54) | 15.65 (7.33, 28.00) | 19.95 (11.67, 27.24) |
|  |  | High | 79.76 (59.62, 100.77) | 80.31 (52.17, 112.17) | 34.60 (27.42, 41.39) |
|  |  | Very High | 315.41 (261.44, 377.24) | 313.91 (243.04, 384.24) | 62.77 (57.81, 67.39) |
|  | 4 | Low | NA | NA | NA |
|  |  | Medium | NA | NA | NA |
|  |  | High | 79.63 (72.91, 84.81) | 75.01 (63.92, 82.75) | 93.89 (91.82, 95.49) |
|  |  | Very High | 60.78 (45.49, 71.86) | 51.75 (30.91, 68.06) | 97.36 (96.32, 98.14) |

## Appendix B

The tables below present the median and first and third quartiles of the percent error calculated across each combination of inter-study variance and mean skew levels in the primary and sensitivity analyses. The data set was restricted to the scenario where the number of studies was 50 and the median number of subjects per study was 100. Studies with absolute percent error greater than 500% were removed. The format for the entries of the following table is: median (first quartile, third quartile). Entries with NA correspond to scenarios that were not observed in the simulation study.

**Table B1.** Median-based approaches.

| Approach | $\tau^2$ | Skew | Medians Given | Mix Given | Means Given |
|---|---|---|---|---|---|
| MM | 1/16 | Low | 0.03 (-2.79, 3.09) | 0.62 (-2.29, 3.71) | 3.17 (0.11, 6.19) |
| | | Medium | -0.16 (-3.11, 2.93) | 0.14 (-2.84, 3.22) | 11.96 (7.81, 15.81) |
| | | High | 0.04 (-3.21, 3.40) | 0.07 (-3.19, 3.40) | 63.00 (57.19, 68.57) |
| | | Very High | 0.26 (-3.82, 4.34) | 0.26 (-3.82, 4.34) | 476.16 (459.40, 489.17) |
| | 1/4 | Low | -0.13 (-5.77, 6.33) | 0.52 (-5.22, 6.89) | 3.05 (-2.86, 9.65) |
| | | Medium | 0.13 (-5.73, 5.94) | 0.35 (-5.48, 6.24) | 11.72 (4.88, 18.67) |
| | | High | 0.09 (-6.10, 6.20) | 0.09 (-6.07, 6.23) | 62.73 (52.81, 73.67) |
| | | Very High | 0.42 (-6.37, 6.55) | 0.42 (-6.37, 6.55) | 468.29 (440.65, 485.45) |
| | 1 | Low | 7.46 (1.54, 15.78) | 7.46 (1.54, 15.78) | 19.39 (11.30, 22.87) |
| | | Medium | 0.14 (-11.54, 12.65) | 0.40 (-11.36, 12.91) | 13.24 (0.43, 27.43) |
| | | High | -0.41 (-11.43, 12.83) | -0.41 (-11.39, 12.88) | 61.78 (43.25, 83.28) |
| | | Very High | 0.33 (-11.57, 13.56) | 0.33 (-11.57, 13.56) | 441.53 (398.60, 471.30) |
| | 4 | Low | NA | NA | NA |
| | | Medium | NA | NA | NA |
| | | High | 0.12 (-20.76, 27.36) | 0.15 (-20.75, 27.42) | 63.47 (30.06, 107.97) |
| | | Very High | -0.43 (-21.84, 27.57) | -0.43 (-21.84, 27.57) | 384.45 (316.45, 443.05) |
| WM | 1/16 | Low | 0.10 (-3.63, 4.04) | 0.36 (-3.37, 4.28) | 3.27 (-0.58, 7.18) |
| | | Medium | -0.11 (-3.84, 3.81) | -0.03 (-3.73, 3.94) | 11.87 (7.08, 16.81) |
| | | High | 0.15 (-3.64, 4.05) | 0.15 (-3.63, 4.05) | 63.26 (57.24, 70.20) |
| | | Very High | 0.35 (-4.19, 5.03) | 0.35 (-4.19, 5.03) | 479.15 (466.13, 490.90) |
| | 1/4 | Low | -0.29 (-7.10, 7.87) | -0.02 (-6.96, 8.31) | 2.86 (-4.25, 11.19) |
| | | Medium | 0.13 (-7.02, 7.88) | 0.28 (-6.98, 8.02) | 11.82 (3.46, 20.75) |
| | | High | -0.28 (-7.51, 7.69) | -0.28 (-7.51, 7.71) | 63.14 (50.69, 76.74) |
| | | Very High | -0.29 (-7.51, 8.20) | -0.29 (-7.51, 8.20) | 462.21 (436.70, 482.55) |
| | 1 | Low | 16.99 (1.40, 55.08) | 16.99 (1.40, 55.08) | 34.06 (15.45, 77.13) |
| | | Medium | 0.20 (-13.72, 16.04) | 0.21 (-13.65, 16.07) | 13.25 (-1.94, 31.76) |
| | | High | 0.13 (-14.28, 15.88) | 0.13 (-14.27, 15.88) | 62.17 (39.39, 89.64) |
| | | Very High | 0.03 (-14.51, 17.38) | 0.03 (-14.51, 17.38) | 433.11 (385.05, 469.17) |
| | 4 | Low | NA | NA | NA |

| | | | |
|---|---|---|---|
| Medium | NA | NA | NA |
| High | 0.29 (-25.76, 33.20) | 0.29 (-25.76, 33.20) | 63.70 (22.19, 117.78) |
| Very High | -0.91 (-26.90, 33.30) | -0.91 (-26.90, 33.30) | 362.54 (279.24, 431.46) |

**Table B2.** Transformation-based approaches. In the column "Means Given", we present the percent error of the MEANS$_{FE}$ approach for the fixed effect transformation-based methods and the percent error of the MEANS$_{RE}$ approach for the random effects transformation-based methods.

| Approach | $\tau^2$ | Skew | Medians Given | Mix Given | Means Given |
|---|---|---|---|---|---|
| T1$_{FE}$ | 1/16 | Low | -13.19 (-16.08, -10.42) | -12.73 (-16.17, -8.90) | -11.33 (-14.27, -8.50) |
| | | Medium | -18.44 (-21.54, -15.35) | -17.91 (-21.78, -13.96) | -11.85 (-14.88, -8.94) |
| | | High | -38.87 (-41.10, -36.66) | -38.56 (-41.55, -35.14) | -14.95 (-18.02, -11.95) |
| | | Very High | -81.19 (-81.98, -80.38) | -81.07 (-82.19, -79.95) | -43.07 (-47.65, -38.88) |
| | 1/4 | Low | -38.81 (-44.17, -33.80) | -37.90 (-44.35, -31.57) | -37.60 (-42.84, -32.42) |
| | | Medium | -42.24 (-47.51, -37.26) | -40.73 (-47.29, -34.45) | -37.68 (-43.23, -32.48) |
| | | High | -56.83 (-60.64, -53.22) | -55.99 (-60.98, -50.70) | -40.02 (-45.39, -35.00) |
| | | Very High | -86.73 (-87.94, -85.58) | -86.48 (-88.14, -84.81) | -59.45 (-64.64, -54.33) |
| | 1 | Low | -81.42 (-82.61, -79.37) | -77.31 (-82.71, -74.31) | -79.90 (-81.15, -76.41) |
| | | Medium | -83.74 (-87.81, -79.62) | -82.10 (-87.08, -76.38) | -82.33 (-86.80, -77.91) |
| | | High | -87.77 (-90.78, -84.59) | -86.38 (-90.29, -82.04) | -83.08 (-87.18, -78.64) |
| | | Very High | -96.21 (-97.16, -95.25) | -95.75 (-96.96, -94.35) | -87.92 (-91.23, -84.25) |
| | 4 | Low | NA | NA | NA |
| | | Medium | NA | NA | NA |
| | | High | -99.75 (-99.88, -99.53) | -99.61 (-99.83, -99.21) | -99.64 (-99.83, -99.34) |
| | | Very High | -99.92 (-99.96, -99.85) | -99.88 (-99.95, -99.76) | -99.73 (-99.87, -99.49) |
| T2$_{FE}$ | 1/16 | Low | -4.54 (-7.82, -1.29) | -5.74 (-9.67, -1.82) | -11.33 (-14.27, -8.50) |
| | | Medium | 13.28 (7.07, 18.10) | 12.28 (4.15, 18.99) | -11.85 (-14.88, -8.94) |
| | | High | 80.50 (70.03, 90.93) | 81.44 (66.18, 96.62) | -14.95 (-18.02, -11.95) |
| | | Very High | 127.92 (92.23, 163.43) | 134.48 (87.77, 189.80) | -43.07 (-47.65, -38.88) |
| | 1/4 | Low | -32.76 (-38.41, -27.20) | -32.96 (-39.61, -26.11) | -37.60 (-42.84, -32.42) |
| | | Medium | -21.16 (-29.14, -13.50) | -20.70 (-30.02, -11.12) | -37.68 (-43.23, -32.48) |
| | | High | 27.07 (13.33, 40.32) | 30.28 (12.48, 48.26) | -40.02 (-45.39, -35.00) |
| | | Very High | 65.55 (32.03, 101.10) | 75.04 (32.53, 124.27) | -59.45 (-64.64, -54.33) |
| | 1 | Low | -72.06 (-76.50, -69.25) | -71.11 (-72.99, -70.58) | -79.90 (-81.15, -76.41) |
| | | Medium | -76.98 (-82.76, -71.24) | -74.86 (-81.82, -66.72) | -82.33 (-86.80, -77.91) |
| | | High | -64.13 (-73.28, -53.85) | -60.37 (-71.92, -45.74) | -83.08 (-87.18, -78.64) |
| | | Very High | -48.24 (-65.23, -28.90) | -40.08 (-62.08, -10.22) | -87.92 (-91.23, -84.25) |
| | 4 | Low | NA | NA | NA |
| | | Medium | NA | NA | NA |
| | | High | -99.23 (-99.64, -98.57) | -98.83 (-99.48, -97.69) | -99.64 (-99.83, -99.34) |
| | | Very High | -98.73 (-99.43, -97.48) | -97.96 (-99.14, -95.61) | -99.73 (-99.87, -99.49) |
| T1$_{RE}$ | 1/16 | Low | -2.49 (-4.71, -0.01) | -1.94 (-4.73, 1.11) | -0.36 (-2.63, 2.15) |
| | | Medium | -8.44 (-10.96, -5.61) | -7.95 (-11.32, -4.31) | -1.05 (-3.44, 1.52) |
| | | High | -31.24 (-33.01, -29.16) | -31.19 (-33.78, -28.26) | -4.36 (-6.87, -1.70) |
| | | Very High | -78.16 (-78.85, -77.43) | -78.12 (-79.16, -77.07) | -28.79 (-31.87, -25.52) |
| | 1/4 | Low | -2.91 (-7.87, 2.28) | -2.44 (-8.69, 4.01) | -0.80 (-5.91, 4.42) |
| | | Medium | -9.29 (-13.96, -4.07) | -8.74 (-15.15, -2.10) | -2.23 (-6.95, 2.94) |
| | | High | -33.24 (-36.73, -29.61) | -33.37 (-38.17, -28.13) | -8.14 (-13.02, -3.11) |
| | | Very High | -79.34 (-80.50, -78.20) | -79.40 (-80.95, -77.71) | -35.61 (-39.64, -30.80) |
| | 1 | Low | -14.40 (-18.10, -10.46) | -14.02 (-20.23, -7.04) | -8.82 (-12.57, -4.05) |
| | | Medium | -25.44 (-32.42, -17.06) | -25.25 (-34.96, -14.13) | -19.77 (-27.21, -11.05) |
| | | High | -50.15 (-55.11, -44.62) | -49.27 (-56.37, -40.96) | -34.60 (-41.39, -27.42) |
| | | Very High | -86.24 (-87.71, -84.60) | -85.78 (-87.93, -83.47) | -62.77 (-67.39, -57.81) |
| | 4 | Low | NA | NA | NA |
| | | Medium | NA | NA | NA |
| | | High | -94.87 (-96.18, -93.19) | -93.69 (-95.70, -91.03) | -93.89 (-95.49, -91.82) |
| | | Very High | -98.82 (-99.14, -98.40) | -98.54 (-99.02, -97.82) | -97.36 (-98.14, -96.32) |
| $T2_{RE}$ | 1/16 | Low | 7.05 (4.50, 9.71) | 4.64 (1.42, 7.88) | -0.36 (-2.63, 2.15) |
| | | Medium | 28.54 (23.12, 32.64) | 26.20 (18.91, 32.46) | -1.05 (-3.44, 1.52) |
| | | High | 138.05 (125.67, 150.75) | 136.66 (119.41, 156.28) | -4.36 (-6.87, -1.70) |
| | | Very High | 466.09 (433.28, 483.46) | 439.24 (391.85, 472.36) | -28.79 (-31.87, -25.52) |
| | 1/4 | Low | 6.57 (1.00, 12.29) | 3.97 (-2.44, 10.81) | -0.80 (-5.91, 4.42) |
| | | Medium | 26.02 (17.42, 33.62) | 23.62 (13.07, 33.78) | -2.23 (-6.95, 2.94) |
| | | High | 132.70 (115.38, 149.19) | 131.23 (108.51, 156.26) | -8.14 (-13.02, -3.11) |
| | | Very High | 450.02 (411.75, 478.19) | 420.10 (365.88, 461.99) | -35.61 (-39.64, -30.80) |
| | 1 | Low | 23.13 (19.46, 27.04) | 17.39 (15.29, 27.76) | -8.82 (-12.57, -4.05) |
| | | Medium | 8.77 (-1.66, 20.60) | 7.59 (-6.96, 25.22) | -19.77 (-27.21, -11.05) |
| | | High | 79.76 (59.62, 100.77) | 80.31 (52.17, 112.17) | -34.60 (-41.39, -27.42) |
| | | Very High | 315.41 (261.44, 377.24) | 313.91 (243.04, 384.24) | -62.77 (-67.39, -57.81) |
| | 4 | Low | NA | NA | NA |
| | | Medium | NA | NA | NA |
| | | High | -79.63 (-84.81, -72.91) | -75.01 (-82.75, -63.92) | -93.89 (-95.49, -91.82) |
| | | Very High | -60.75 (-71.85, -45.43) | -49.31 (-66.83, -24.90) | -97.36 (-98.14, -96.32) |

## Appendix C

The MSE was calculated over the 1,000 generated data sets for each combination of data generation parameters in the simulation study. The tables below present the median and first and third quartiles of the MSE calculated across each combination of inter-study variance and mean skew levels in the primary and sensitivity analyses. The data set was restricted to the scenario where the number of studies was 50 and the median number of subjects per study was 100. Studies with absolute percent error greater than 500% were removed. The format for the entries of the following table is: median (first quartile, third quartile). Entries with NA correspond to scenarios that were not observed in the simulation study.

**Table C1.** Median-based approaches.

| Approach | $\tau^2$ | Skew | Medians Given | Mix Given | Means Given |
|---|---|---|---|---|---|
| MM | 1/16 | Low | 0.10 (0.05, 0.16) | 0.05 (0.05, 0.16) | 0.13 (0.07, 0.19) |
|  |  | Medium | 0.12 (0.04, 0.16) | 0.06 (0.04, 0.15) | 0.50 (0.37, 0.65) |
|  |  | High | 0.07 (0.03, 0.08) | 0.07 (0.03, 0.11) | 3.70 (3.51, 9.83) |
|  |  | Very High | 0.01 (0.00, 0.13) | 0.01 (0.00, 0.19) | 8.78 (8.49, 442.20) |
|  | 1/4 | Low | 0.47 (0.15, 0.62) | 0.21 (0.16, 0.54) | 0.53 (0.19, 0.70) |
|  |  | Medium | 0.30 (0.13, 0.67) | 0.20 (0.14, 0.43) | 0.77 (0.44, 1.34) |
|  |  | High | 0.20 (0.07, 0.63) | 0.20 (0.07, 0.24) | 3.64 (3.07, 12.14) |
|  |  | Very High | 0.01 (0.00, 0.30) | 0.01 (0.00, 0.43) | 7.02 (6.47, 411.17) |
|  | 1 | Low | NA | NA | NA |
|  |  | Medium | 0.83 (0.26, 2.73) | 0.84 (0.26, 0.92) | 1.38 (0.49, 4.28) |
|  |  | High | 0.46 (0.12, 3.00) | 0.46 (0.12, 0.87) | 3.00 (1.83, 20.52) |
|  |  | Very High | 0.02 (0.01, 3.23) | 0.02 (0.01, 0.94) | 3.18 (2.92, 365.26) |
|  | 4 | Low | NA | NA | NA |
|  |  | Medium | NA | NA | NA |
|  |  | High | 0.12 (0.03, 15.65) | 0.12 (0.03, 3.94) | 0.35 (0.16, 43.05) |
|  |  | Very High | 0.01 (0.00, 17.97) | 0.01 (0.00, 4.58) | 0.12 (0.11, 250.22) |
| WM | 1/16 | Low | 0.13 (0.06, 0.24) | 0.08 (0.06, 0.24) | 0.17 (0.09, 0.28) |
|  |  | Medium | 0.16 (0.06, 0.22) | 0.09 (0.06, 0.21) | 0.54 (0.41, 0.72) |
|  |  | High | 0.09 (0.03, 0.11) | 0.09 (0.03, 0.15) | 3.87 (3.61, 10.02) |
|  |  | Very High | 0.01 (0.00, 0.14) | 0.01 (0.00, 0.22) | 8.87 (8.57, 444.70) |
|  | 1/4 | Low | 0.58 (0.22, 0.78) | 0.31 (0.23, 0.79) | 0.65 (0.28, 0.90) |
|  |  | Medium | 0.41 (0.21, 0.79) | 0.32 (0.21, 0.72) | 0.94 (0.55, 1.47) |
|  |  | High | 0.31 (0.10, 0.88) | 0.31 (0.10, 0.36) | 4.22 (3.34, 12.80) |
|  |  | Very High | 0.02 (0.01, 0.33) | 0.02 (0.01, 0.56) | 7.07 (6.21, 410.50) |
|  | 1 | Low | NA | NA | NA |
|  |  | Medium | 1.37 (0.39, 3.62) | 1.37 (0.39, 1.78) | 2.36 (0.69, 5.62) |
|  |  | High | 0.83 (0.19, 4.20) | 0.83 (0.19, 1.51) | 3.91 (2.13, 24.29) |
|  |  | Very High | 0.04 (0.01, 4.52) | 0.04 (0.01, 1.59) | 3.05 (2.93, 339.35) |

| | | | | |
|---|---|---|---|---|
| 4 | Low | NA | NA | NA |
| | Medium | NA | NA | NA |
| | High | 0.19 (0.05, 18.65) | 0.19 (0.05, 7.51) | 0.41 (0.23, 50.90) |
| | Very High | 0.01 (0.00, 21.64) | 0.01 (0.00, 7.52) | 0.11 (0.11, 229.31) |

**Table C2.** Transformation-based approaches. In the column "Means Given", we present the MSE of the MEANS$_{FE}$ approach for the fixed effect transformation-based methods and the MSE of the MEANS$_{RE}$ approach for the random effects transformation-based methods.

| Approach | $\tau^2$ | Skew | Medians Given | Mix Given | Means Given |
|---|---|---|---|---|---|
| T1$_{FE}$ | 1/16 | Low | 0.55 (0.52, 0.58) | 0.54 (0.50, 0.57) | 0.43 (0.40, 0.46) |
| | | Medium | 1.32 (0.99, 1.36) | 1.28 (0.98, 1.32) | 0.56 (0.45, 0.63) |
| | | High | 4.20 (3.96, 11.42) | 4.14 (3.89, 11.18) | 1.00 (0.83, 2.39) |
| | | Very High | 17.34 (16.77, 975.94) | 17.24 (16.61, 966.76) | 7.71 (5.84, 347.32) |
| | 1/4 | Low | 5.00 (3.94, 5.27) | 4.69 (3.81, 5.03) | 4.71 (3.69, 4.96) |
| | | Medium | 7.14 (4.81, 7.80) | 6.65 (4.58, 7.24) | 5.64 (3.83, 6.22) |
| | | High | 26.59 (8.23, 28.73) | 24.75 (8.04, 27.12) | 13.06 (4.27, 14.98) |
| | | Very High | 19.44 (18.94, 1326.39) | 19.28 (18.62, 1301.71) | 11.67 (9.72, 675.07) |
| | 1 | Low | NA | NA | NA |
| | | Medium | 54.95 (17.48, 59.05) | 48.03 (16.70, 54.82) | 52.43 (16.91, 56.97) |
| | | High | 131.67 (19.15, 138.31) | 119.14 (18.42, 130.02) | 113.70 (17.11, 122.50) |
| | | Very High | 3341.41 (23.13, 3402.19) | 3231.99 (22.81, 3330.41) | 2582.80 (19.28, 2799.29) |
| | 4 | Low | NA | NA | NA |
| | | Medium | NA | NA | NA |
| | | High | 3647.84 (24.83, 3667.77) | 3568.26 (24.71, 3618.19) | 3622.73 (24.77, 3650.10) |
| | | Very High | 74105.69 (24.95, 74250.93) | 73580.04 (24.91, 73950.56) | 72986.37 (24.81, 73605.09) |
| T2$_{FE}$ | 1/16 | Low | 0.22 (0.16, 0.24) | 0.33 (0.25, 0.37) | 0.43 (0.40, 0.46) |
| | | Medium | 0.64 (0.18, 0.90) | 0.71 (0.31, 1.12) | 0.56 (0.45, 0.63) |
| | | High | 10.69 (5.47, 19.75) | 18.28 (7.18, 24.87) | 1.00 (0.83, 2.39) |
| | | Very High | 89.64 (28.85, 501.33) | 125.27 (43.68, 1237.79) | 7.71 (5.84, 347.32) |
| | 1/4 | Low | 3.71 (2.99, 4.04) | 3.74 (3.14, 4.21) | 4.71 (3.69, 4.96) |
| | | Medium | 2.12 (1.67, 2.83) | 2.34 (2.00, 3.34) | 5.64 (3.83, 6.22) |
| | | High | 5.35 (2.42, 8.30) | 10.82 (3.70, 13.49) | 13.06 (4.27, 14.98) |
| | | Very High | 47.37 (10.94, 329.70) | 84.60 (21.30, 979.73) | 11.67 (9.72, 675.07) |
| | 1 | Low | NA | NA | NA |
| | | Medium | 44.06 (14.85, 50.81) | 39.00 (13.88, 46.35) | 52.43 (16.91, 56.97) |
| | | High | 59.39 (10.72, 79.30) | 58.27 (9.43, 69.50) | 113.70 (17.11, 122.50) |
| | | Very High | 998.79 (9.07, 1432.08) | 1035.94 (11.61, 1714.36) | 2582.80 (19.28, 2799.29) |
| | 4 | Low | NA | NA | NA |
| | | Medium | NA | NA | NA |
| | | High | 3516.01 (24.49, 3597.88) | 3310.81 (24.11, 3484.07) | 3622.73 (24.77, 3650.10) |
| | | Very High | 66943.31 (24.09, 71133.39) | 59764.66 (23.38, 67490.46) | 72986.37 (24.81, 73605.09) |
| $T1_{RE}$ | 1/16 | Low | 0.09 (0.05, 0.13) | 0.06 (0.06, 0.18) | 0.08 (0.03, 0.12) |
| | | Medium | 0.31 (0.25, 0.39) | 0.33 (0.26, 0.41) | 0.11 (0.04, 0.14) |
| | | High | 2.67 (2.53, 7.38) | 2.67 (2.62, 7.59) | 0.23 (0.18, 0.51) |
| | | Very High | 15.66 (15.27, 884.26) | 15.65 (15.25, 890.84) | 4.22 (2.18, 124.15) |
| | 1/4 | Low | 0.45 (0.16, 0.58) | 0.30 (0.22, 0.81) | 0.45 (0.14, 0.59) |
| | | Medium | 0.62 (0.41, 1.03) | 0.78 (0.45, 1.01) | 0.36 (0.16, 0.73) |
| | | High | 3.25 (2.91, 10.69) | 3.37 (3.12, 11.11) | 0.76 (0.54, 2.02) |
| | | Very High | 16.29 (15.71, 1097.61) | 16.32 (15.73, 1107.38) | 5.58 (3.22, 225.05) |
| | 1 | Low | NA | NA | NA |
| | | Medium | 2.80 (2.44, 8.60) | 4.35 (2.56, 9.12) | 2.28 (1.92, 6.79) |
| | | High | 7.98 (6.36, 50.81) | 7.66 (6.69, 47.67) | 4.88 (3.31, 28.83) |
| | | Very High | 19.55 (18.54, 2791.90) | 19.37 (18.31, 2720.13) | 12.56 (9.85, 1577.67) |
| | 4 | Low | NA | NA | NA |
| | | Medium | NA | NA | NA |
| | | High | 23.01 (22.36, 3233.79) | 22.39 (21.61, 3011.04) | 22.56 (21.85, 3125.83) |
| | | Very High | 24.56 (24.36, 72212.23) | 24.41 (24.17, 70669.79) | 23.96 (23.57, 68867.59) |
| $T2_{RE}$ | 1/16 | Low | 0.18 (0.11, 0.25) | 0.13 (0.07, 0.25) | 0.08 (0.03, 0.12) |
| | | Medium | 2.25 (1.37, 2.57) | 2.11 (1.15, 2.75) | 0.11 (0.04, 0.14) |
| | | High | 49.99 (18.78, 52.67) | 51.66 (21.93, 59.05) | 0.23 (0.18, 0.51) |
| | | Very High | 428.07 (165.03, 8421.28) | 458.87 (168.07, 9651.48) | 4.22 (2.18, 124.15) |
| | 1/4 | Low | 0.56 (0.28, 0.75) | 0.40 (0.26, 0.88) | 0.45 (0.14, 0.59) |
| | | Medium | 2.36 (1.60, 3.03) | 3.12 (1.48, 3.82) | 0.36 (0.16, 0.73) |
| | | High | 50.98 (45.68, 59.84) | 55.62 (47.95, 79.70) | 0.76 (0.54, 2.02) |
| | | Very High | 393.93 (125.13, 7601.29) | 429.24 (143.68, 9552.92) | 5.58 (3.22, 225.05) |
| | 1 | Low | NA | NA | NA |
| | | Medium | 2.98 (0.98, 6.28) | 5.95 (1.82, 7.39) | 2.28 (1.92, 6.79) |
| | | High | 30.13 (14.81, 46.52) | 45.09 (23.55, 111.12) | 4.88 (3.31, 28.83) |
| | | Very High | 273.67 (247.80, 6427.09) | 263.73 (201.36, 10591.90) | 12.56 (9.85, 1577.67) |
| | 4 | Low | NA | NA | NA |
| | | Medium | NA | NA | NA |
| | | High | 19.48 (15.58, 2370.76) | 18.13 (13.62, 2106.78) | 22.56 (21.85, 3125.83) |
| | | Very High | 18.29 (9.97, 37812.04) | 25.88 (16.02, 47666.76) | 23.96 (23.57, 68867.59) |

**Figure C1.** Interaction plots of the primary and sensitivity analyses of the approach by the mean skew level. Performance is measured by MSE.

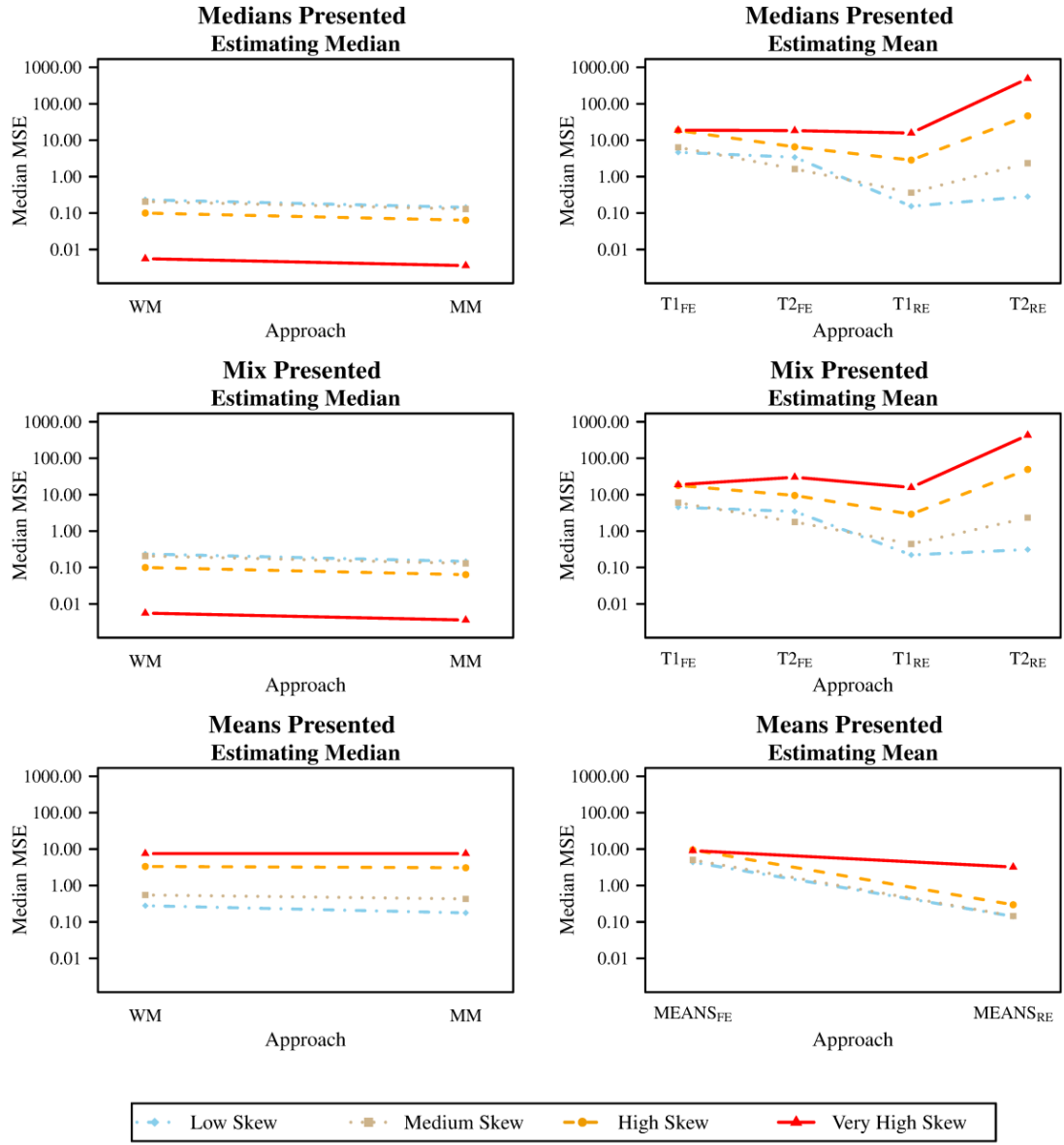

The MSE was calculated over the 1,000 generated data sets for each combination of data generation parameters. The data set was restricted to the scenario where number of studies equalled 50, the median number of subjects per study equalled 100, and inter-study variance equalled 1/4. Studies with APE greater than 500% were removed. The y-coordinates are plotted on a log scale in all plots in this figure.

**Figure C2.** Interaction plots of the primary and sensitivity analyses of the approach by the inter-study variance. Performance is measured by MSE.

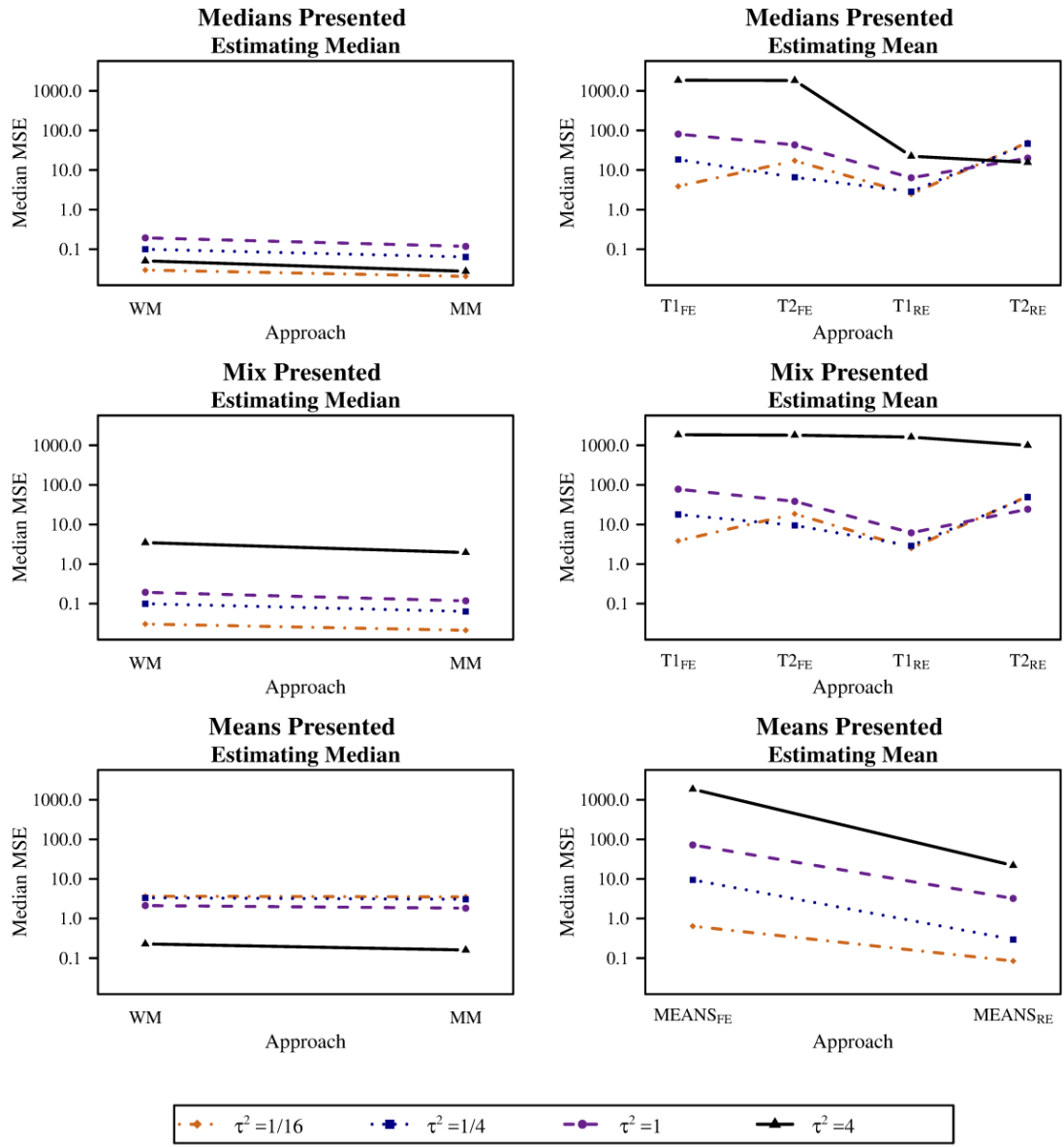

The MSE was calculated over the 1,000 generated data sets for each combination of data generation parameters. The data set was restricted to the scenario where number of studies equalled 50, the median number of subjects per study equalled 100, and the mean skew level was classified as high ($0.2 < $ mean $SK_b \leq 0.4$). Studies with APE greater than 500% were removed. The y-coordinates are plotted on a log scale in all plots in this figure.

## Appendix D

The tables below present the coverage of the 95% CIs calculated across each combination of inter-study variance and mean skew levels in the primary and sensitivity analyses. The data set was restricted to the scenario where the number of studies was 50 and the median number of subjects per study was 100. Studies with absolute percent error greater than 500% were removed. Entries with NA correspond to scenarios that were not observed in the simulation study.

**Table D1.** Median-based approaches.

| Approach | $\tau^2$ | Skew | Medians Given | Mix Given | Means Given |
|---|---|---|---|---|---|
| MM | 1/16 | Low | 0.95 | 0.95 | 0.89 |
| | | Medium | 0.94 | 0.94 | 0.33 |
| | | High | 0.95 | 0.95 | 0.01 |
| | | Very High | 0.95 | 0.95 | 0.00 |
| | 1/4 | Low | 0.95 | 0.95 | 0.93 |
| | | Medium | 0.95 | 0.95 | 0.76 |
| | | High | 0.94 | 0.94 | 0.02 |
| | | Very High | 0.95 | 0.95 | 0.00 |
| | 1 | Low | 1.00 | 1.00 | 1.00 |
| | | Medium | 0.95 | 0.95 | 0.88 |
| | | High | 0.95 | 0.95 | 0.23 |
| | | Very High | 0.95 | 0.95 | 0.00 |
| | 4 | Low | NA | NA | NA |
| | | Medium | NA | NA | NA |
| | | High | 0.95 | 0.95 | 0.71 |
| | | Very High | 0.95 | 0.95 | 0.00 |
| WM | 1/16 | Low | 0.88 | 0.88 | 0.82 |
| | | Medium | 0.87 | 0.87 | 0.36 |
| | | High | 0.88 | 0.88 | 0.01 |
| | | Very High | 0.88 | 0.88 | 0.00 |
| | 1/4 | Low | 0.88 | 0.88 | 0.87 |
| | | Medium | 0.89 | 0.89 | 0.71 |
| | | High | 0.88 | 0.88 | 0.02 |
| | | Very High | 0.88 | 0.88 | 0.00 |
| | 1 | Low | 0.83 | 0.83 | 0.83 |
| | | Medium | 0.89 | 0.89 | 0.83 |
| | | High | 0.89 | 0.89 | 0.29 |
| | | Very High | 0.88 | 0.88 | 0.00 |
| | 4 | Low | NA | NA | NA |
| | | Medium | NA | NA | NA |
| | | High | 0.88 | 0.88 | 0.67 |
| | | Very High | 0.89 | 0.89 | 0.01 |

**Table D2.** Transformation-based approaches. In the column "Means Given", we present the coverage of the 95% CIs of the MEANS$_{FE}$ approach for the fixed effect transformation-based methods and the coverage of the 95% CIs of the MEANS$_{RE}$ for the random effects transformation-based methods.

| Approach | $\tau^2$ | Skew | Medians Given | Mix Given | Means Given |
|---|---|---|---|---|---|
| T1$_{FE}$ | 1/16 | Low | 0.00 | 0.01 | 0.00 |
| | | Medium | 0.00 | 0.00 | 0.01 |
| | | High | 0.00 | 0.00 | 0.00 |
| | | Very High | 0.00 | 0.00 | 0.00 |
| | 1/4 | Low | 0.00 | 0.00 | 0.00 |
| | | Medium | 0.00 | 0.00 | 0.00 |
| | | High | 0.00 | 0.00 | 0.00 |
| | | Very High | 0.00 | 0.00 | 0.00 |
| | 1 | Low | 0.00 | 0.00 | 0.00 |
| | | Medium | 0.00 | 0.00 | 0.00 |
| | | High | 0.00 | 0.00 | 0.00 |
| | | Very High | 0.00 | 0.00 | 0.00 |
| | 4 | Low | NA | NA | NA |
| | | Medium | NA | NA | NA |
| | | High | 0.00 | 0.00 | 0.00 |
| | | Very High | 0.00 | 0.00 | 0.00 |
| | | | | | |
| T2$_{FE}$ | 1/16 | Low | 0.06 | 0.07 | 0.00 |
| | | Medium | 0.02 | 0.04 | 0.01 |
| | | High | 0.00 | 0.00 | 0.00 |
| | | Very High | 0.00 | 0.01 | 0.00 |
| | 1/4 | Low | 0.00 | 0.00 | 0.00 |
| | | Medium | 0.01 | 0.03 | 0.00 |
| | | High | 0.03 | 0.04 | 0.00 |
| | | Very High | 0.02 | 0.03 | 0.00 |
| | 1 | Low | 0.00 | 0.00 | 0.00 |
| | | Medium | 0.00 | 0.00 | 0.00 |
| | | High | 0.00 | 0.01 | 0.00 |
| | | Very High | 0.02 | 0.04 | 0.00 |
| | 4 | Low | NA | NA | NA |
| | | Medium | NA | NA | NA |
| | | High | 0.00 | 0.00 | 0.00 |
| | | Very High | 0.00 | 0.00 | 0.00 |
| | | | | | |
| T1$_{RE}$ | 1/16 | Low | 0.84 | 0.87 | 0.92 |
|  |  | Medium |  | 0.28 | 0.51 | 0.89 |
|  |  | High |  | 0.00 | 0.01 | 0.73 |
|  |  | Very High |  | 0.00 | 0.00 | 0.00 |
|  | 1/4 | Low |  | 0.72 | 0.74 | 0.76 |
|  |  | Medium |  | 0.45 | 0.59 | 0.75 |
|  |  | High |  | 0.01 | 0.02 | 0.56 |
|  |  | Very High |  | 0.00 | 0.00 | 0.00 |
|  | 1 | Low |  | 0.00 | 0.50 | 0.50 |
|  |  | Medium |  | 0.07 | 0.15 | 0.14 |
|  |  | High |  | 0.00 | 0.01 | 0.02 |
|  |  | Very High |  | 0.00 | 0.00 | 0.00 |
|  | 4 | Low | NA | NA | NA |
|  |  | Medium | NA | NA | NA |
|  |  | High |  | 0.00 | 0.00 | 0.00 |
|  |  | Very High |  | 0.00 | 0.00 | 0.00 |
| $T2_{RE}$ | 1/16 | Low |  | 0.47 | 0.80 | 0.92 |
|  |  | Medium |  | 0.08 | 0.13 | 0.89 |
|  |  | High |  | 0.00 | 0.00 | 0.73 |
|  |  | Very High |  | 0.00 | 0.00 | 0.00 |
|  | 1/4 | Low |  | 0.65 | 0.76 | 0.76 |
|  |  | Medium |  | 0.12 | 0.29 | 0.75 |
|  |  | High |  | 0.00 | 0.00 | 0.56 |
|  |  | Very High |  | 0.00 | 0.00 | 0.00 |
|  | 1 | Low |  | 0.17 | 0.17 | 0.50 |
|  |  | Medium |  | 0.32 | 0.38 | 0.14 |
|  |  | High |  | 0.01 | 0.03 | 0.02 |
|  |  | Very High |  | 0.00 | 0.00 | 0.00 |
|  | 4 | Low | NA | NA | NA |
|  |  | Medium | NA | NA | NA |
|  |  | High |  | 0.00 | 0.01 | 0.00 |
|  |  | Very High |  | 0.01 | 0.07 | 0.00 |